\documentclass[journal,letter,10pt,twocolumn]{IEEEtran}
\pdfoutput=1

\usepackage{amsmath, amssymb, amscd, amsthm, amsfonts}
\usepackage{bbm}
\usepackage[pdftex]{graphicx}
\usepackage[caption=false,font=footnotesize]{subfig}
\usepackage{dblfloatfix}    
\usepackage{color}
\usepackage{algorithm}
\usepackage[noend]{algpseudocode}
\usepackage{comment}

\usepackage{hyperref}

\usepackage{tikz,ifthen}
\usetikzlibrary{positioning,arrows,backgrounds}
\usetikzlibrary{fit}
\usetikzlibrary{calc}
 \usepackage{color}
 \usepackage{transparent}

\newtheorem{theorem}{Theorem}
\newtheorem{lemma}{Lemma}
\newtheorem{corollary}{Corollary}
\newtheorem{proposition}{Proposition}
\newtheorem{definition}{Definition}

\usepackage{pgfplots}
\pgfplotsset{width=10cm,compat=1.9}

\usepackage{subfiles}
\usepackage[noadjust]{cite}

\usepackage[english]{babel}

\addto\captionsenglish{}

\graphicspath{{../Figures/}}



\begin{document}

\title{On the Capacity-Achieving Input~of~the Gaussian Channel with Polar Quantization}

\author{
   \IEEEauthorblockN{Neil Irwin Bernardo, \textit{Graduate Student Member, IEEE}, Jingge Zhu, \textit{Member, IEEE}, and Jamie Evans, \textit{Senior Member, IEEE}
   }
   \thanks{Manuscript received Jan 20, 2022; revised May 11, 2022; accepted Jul 1, 2022. The work was supported in part by Australian Research Council under project DE210101497. N.I. Bernardo acknowledges the Melbourne Research Scholarship of the University of Melbourne and the DOST-ERDT Faculty Development Fund of the Republic of the Philippines for sponsoring his doctoral studies. The associate editor coordinating the review of this article and approving it for publication was N. Liu. (\emph{Corresponding author: Neil Irwin Bernardo}.) }
   \thanks{N.I. Bernardo is with the Department of Electrical and Electronic Engineering, The University of Melbourne, Parkville, VIC 3010, Australia and also with the Electrical and Electronics Engineering Institute, University of the Philippines Diliman, Quezon City 1101, Philippines (e-mail: bernardon@student.unimelb.edu.au).}
    \thanks{J. Zhu and J. Evans are with the Department of Electrical and Electronic Engineering, The University of Melbourne, Parkville, VIC 3010, Australia (e-mail: jingge.zhu@unimelb.edu.au;
jse@unimelb.edu.au).}
}

\maketitle
\begin{abstract}
The polar receiver architecture is a receiver design that captures the envelope and phase information of the signal rather than its in-phase and quadrature components. Several studies have demonstrated the robustness of polar receivers to phase noise and other nonlinearities. Yet, the information-theoretic limits of polar receivers with finite-precision quantizers have not been investigated in the literature. The main contribution of this work is to identify the optimal signaling strategy for the additive white Gaussian noise (AWGN) channel with polar quantization at the output. More precisely, we show that the capacity-achieving modulation scheme has an amplitude phase shift keying (APSK) structure. Using this result, the capacity of the AWGN channel with polar quantization at the output is established by numerically optimizing the probability mass function of the amplitude. The capacity of the polar-quantized AWGN channel with $b_1$-bit phase quantizer and optimized single-bit magnitude quantizer is also presented. Our numerical findings suggest the existence of signal-to-noise ratio (SNR) thresholds, above which the number of amplitude levels of the optimal APSK scheme and their respective probabilities change abruptly. Moreover, the manner in which the capacity-achieving input evolves with increasing SNR depends on the number of phase quantization bits.  
\end{abstract}

\begin{IEEEkeywords}
Low-resolution ADCs, Capacity, Polar Quantization, Amplitude Phase Shift Keying, AWGN
\end{IEEEkeywords}

\IEEEpeerreviewmaketitle

\section{Introduction}\label{section-intro}

\IEEEPARstart{T}{he} use of low-resolution analog-to-digital converters (ADCs) is seen as an innovative approach to address practical issues in 5G such as massive data processing, high power consumption, and cost \cite{Liu:2019}. In
fact, the cost reduction and energy savings of using low-resolution ADCs readily extend to 6G communications since the novel solutions being developed to meet the 6G performance targets are also power hungry and expensive \cite{Halbauer:2021}. However, this low-power design approach imposes a capacity penalty due to severe nonlinear distortion on the received signal. In addition, these nonlinear quantization effects alter the capacity-achieving modulation scheme and so there is a need to revisit the signal construction and coding strategies for communication channels with low-resolution output quantization. 

There is a rich body of work looking at the performance bounds for quantized channels \cite{Mezghani:2012, Orhan:2015, Gayan:2020, bernardo2021sep} as well as the design of practical receiver architectures with low-resolution ADCs \cite{Wang:2019,Choi:2020}. Nevertheless, there is still an important research gap in terms of identifying the exact structure of the capacity-achieving input for quantized channels. A couple of information-theoretic results have identified the capacity-achieving input distribution for various channel models with symmetric 1-bit ADCs. The work of Singh et al. \cite{Singh:2007,Singh:2009} appears to be the first to examine the fundamental limits of communication channels with low-resolution output. Specifically, they proved that binary antipodal signaling is optimal for real additive white Gaussian noise (AWGN) channels with symmetric 1-bit quantization. Extensions of this work to complex-valued channels with 1-bit in-phase and quadrature (I/Q) ADCs showed that Quadrature Phase Shift Keying (QPSK) is capacity-achieving for the complex-valued AWGN channel \cite{Krone:2010}, the coherent/noncoherent Rayleigh fading channel \cite{Krone:2010,Mezghani:2007}, noncoherent Rician channel \cite{Vu:2019}, and the zero-mean Gaussian mixture channel \cite{Rahman:2020,Rahman2:2020}. Moreover, QPSK maintains its optimality for multiple-input single-output (MISO) channels with 1-bit I/Q ADC transmitter side information \cite{Mo:2014} and multiple access Rayleigh channels with 1-bit I/Q ADC \cite{Ranjbar:2019,Ranjbar:2020}. When the 1-bit quantizer is allowed to be asymmetric in the low signal-to-noise ratio (SNR) regime, the proponents of  \cite{Koch:2013} and \cite{Zhang:2015} showed that an on-off keying structure is optimal in the capacity per unit cost sense under an average power constraint. However, the optimal modulation reverts to binary antipodal signaling when a peak power constraint is imposed or the threshold is not allowed to grow unbounded as SNR vanishes.

The characterization of the optimal input distribution is much less tractable for channels with multi-bit I/Q output quantization; even in the point-to-point case. With some guidance from established optimality conditions, existing studies \cite{Singh:2009,Vu:2019} numerically constructed the capacity-achieving input of channels with multi-bit I/Q output quantization. Our previous works \cite{bernardo2021phase,bernardo2021TIT} considered a different yet still practical phase quantization and proved that $2^b$-phase shift keying (PSK) is capacity-achieving for various channel models with $b$-bit phase quantization at the output. However, such a quantization strategy does not exploit all the available degrees of freedom since the information content placed in the magnitude is thrown away.


To address the limitations of phase quantization, we extend our previous work by investigating the capacity-achieving signaling scheme for channels with polar quantization at the output. Aside from quantized observations of the signal's phase, the receiver can also utilize the quantized observations of the signal's magnitude in order to recover the transmitted message. Magnitude quantization is realized using an envelope detector and an ADC. Meanwhile, phase quantization can be implemented efficiently using time-to-digital converters (TDCs) or by quantizing the output of a phase detector. Analytical and measurement results of wireless receivers equipped with polar quantizers have been provided in \cite{Nazari:2014} showing that polar quantization offers a significant boost in signal-to-quantization noise ratio (SQNR) as compared to I/Q quantization under Gaussian signaling. Effectively, this means that polar quantizers would need fewer number of bits (NoBs) to recover the signal as compared to I/Q quantizers. In addition, polar-based receiver implementations exist that work well with amplitude-phase shift keying (APSK) modulation in terms of power efficiency and phase noise/nonlinearity tolerance \cite{Wang:2020}. Despite this, little attention has been given to the capacity of channels with polar quantization at the output. Most theoretical analyses on polar quantization have been more focused on lossy source coding of the source distribution under a mean square error (MSE) criterion \cite{Pearlman:1979,Peric:2002} or symbol error rate (SER) \cite{Jovanovic:2021} criterion. A recent study \cite{Sezer:2021} considered an all-digital multiple-input multiple-output (MIMO) line-of-sight channel and presented numerical results showing that I/Q quantization at the receiver slightly outperforms polar quantization in terms of achievable rate when the quantizers are designed under an equal output probability criterion. However, such quantizer design is not necessarily optimal and no effort is made to optimize the input distribution.

In this paper, we establish the properties of the capacity-achieving input distribution for a point-to-point AWGN channel with polar-quantized output. These properties are then used to simplify the numerical evaluation of the channel capacity. The central contribution of this work is a rigorous proof that the capacity-achieving modulation scheme for an AWGN channel with polar-quantized output should have an APSK structure (Theorem \ref{theorem:AWGN_case}). Furthermore, the angles of the mass points in the optimal constellation are derived analytically. While the proof techniques are similar to those used in channels with phase quantization and I/Q quantization, the application of these techniques to Gaussian channel with polar quantization is new. We also present some numerical findings on polar-quantized AWGN channel with optimized single-bit magnitude quantizer. Specifically, we observe that the number of amplitude levels increases whenever the SNR exceeds a certain threshold. Moreover, in the low SNR regime, the capacity is achieved by a PSK scheme except for the case when the phase quantizer has two quantization bits. In this special case, the capacity-achieving input has an on-off keying structure. These results provide interesting insights about the connection between the capacity-achieving input and SNR.

\section{Problem Formulation and Main Result}\label{section-problem_statement}

\begin{figure*}[t]
    \centering
    \includegraphics[scale = .85]{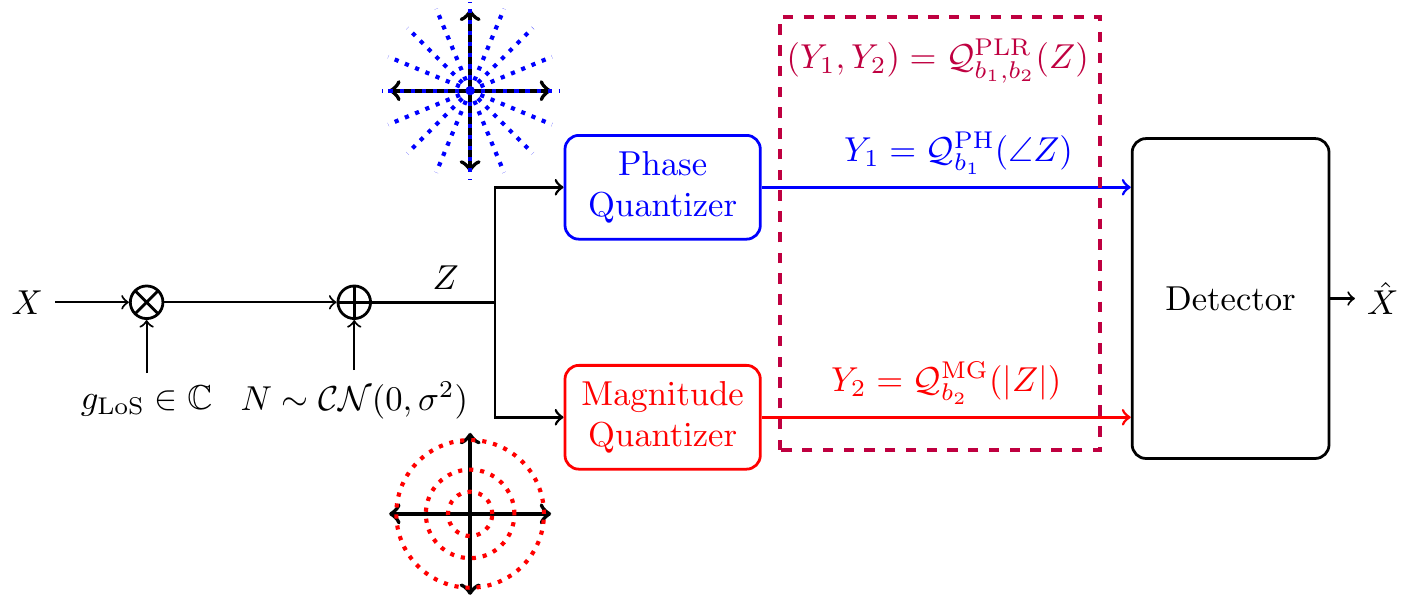}
    \caption{System Model of AWGN Channel with Polar Quantization at the Output}
    \label{fig:polar_receiver_sys_model}
\end{figure*}
We consider a discrete-time\footnote{Synchronized sampling at symbol rate is assumed (i.e. each received sample corresponds to only one transmitted symbol).} memoryless Gaussian channel model with polar quantization at the output as shown in Figure \ref{fig:polar_receiver_sys_model}. The input-output relationship between the transmitted signal $X$ and the unquantized received signal $Z$ is given by
\begin{equation}
    Z = g_{\text{LoS}}X + N,
\end{equation}
where $X$ is the complex input with power constraint $\mathbb{E}\left[|X|^2\right] \leq P$, $N$ is the zero-mean complex Gaussian noise with variance $\sigma^2$, and $g_{\text{LoS}}$ is a complex constant representing the gain and direction of the line-of-sight (LoS) component. The transmitter and receiver are cognizant of the channel gain $g_{\text{LoS}}$. 

The received signal $Z$ is complex-valued and can be represented in polar form as $Z = \sqrt{R}e^{j\Phi}$. The parameters $\Phi$ and $\sqrt{R}$ are then fed to a $b_1$-bit phase quantizer and a $b_2$-bit magnitude quantizers, respectively, to produce the integer-valued outputs, $Y_1$ and $Y_2$. To be more precise, the output of the phase quantizer is the integer $Y_1 = y_1$ if $\Phi \in \mathcal{R}_{y_1}^{\text{PH}}$, where $\mathcal{R}_{y_1}^{\text{PH}}$ is the convex cone given by
\begin{align*}
    \mathcal{R}_{y_1}^{\text{PH}} =  \left\{\phi\in [-\pi,\pi]\;\Big|\;\frac{2\pi}{2^{b_1}}y_1  \leq \phi + \pi < \frac{2\pi(y_1+1)}{2^{b_1}}  \right\},
\end{align*}
and the output of the magnitude quantizer is $Y_2 = y_2$ if $R\in \mathcal{R}_{y_2}^{\text{MG}}$, where $\mathcal{R}_{y_2}^{\text{MG}}$ is given by
\begin{align*}
    \mathcal{R}_{y_2}^{\text{MG}} =  \left\{r\in \mathbb{R}^{+}\;\Big|\;q_{y_2}^2 \leq r < q_{y_2+1}^2 \right\}.
\end{align*}
The quantities $\{q_{l}\}_{l=1}^{l=2^{b_2}-1}$ denote the quantization threshold of the magnitude quantizer and $q_{0} = 0$, and $q_{2^{b_2}} = +\infty$ are implicitly included. Moreover, symmetric phase quantization is considered in this study and is a given in the problem setup\footnote{The symmetric assumption has a practical advantage in time-to-digital converter (TDC)-based implementation of phase quantizers. However, we have not proven the optimality of symmetric phase quantization strategy.}. Due to the circular structure of the phase quantizer branch, the addition operation $Y_1+k$ for some $k\in\mathbb{Z}$ constitutes a modulo $2^{b_1}$ addition. We shall refer to this pair of phase and magnitude quantizers as a $(b_1,b_2)$-bit polar quantizer. The $(b_1,b_2)$-bit polar quantizer is mathematically represented by the function mapping $\mathcal{Q}_{b_1,b_2}^{\mathrm{PLR}}(Z):\;\;\mathbb{C}\mapsto \mathbb{Z}\times\mathbb{Z}$. As a consequence, we shall use the term $(b_1,b_2)$-bit polar-quantized channel to refer to the system model in Figure \ref{fig:polar_receiver_sys_model}. The goal of the detector is to reliably recover the message encoded in $X$ using the quantizer output pair, $(Y_1,Y_2)$. With this problem setup we are now led to the following question: \textit{What should be the distribution of $X$ to maximize the rate of reliable communication when only $(Y_1,Y_2)$ are observed at the receiver side?}

Before presenting our main result, we shall first define a function that will appear frequently in the proofs and main result.

\begin{definition}\label{definition:polar_quantization_func}
Suppose we have $\theta\in[-\pi,\pi]$, $b_1,b_2 \in \mathbb{Z}^{+}$, $y_1\in\{0,\cdots,2^{b_1}-1\}$, $y_2\in\{0,\cdots,2^{b_2}-1\}$, and $\nu \in \mathbb{R}^+$. Suppose further that there is a set $\mathcal{G}_{b_2} = \left\{g_{l}\in\mathbb{R}^+ \big| l\in\{0,\cdots,2^{b_2}\}\right\}$. Then, the polar quantization probability function, denoted as $W_{y_1,y_2}^{(b_1)}(\nu,\theta)$, is defined in \eqref{eq:phase_quantization_prob}. $Q(x)$ is the Gaussian Q function (i.e. the tail probability of a standard normal distribution).
\begin{figure*}[b]
\hrulefill
\begin{equation}\label{eq:phase_quantization_prob}
    W_{y_1,y_2}^{(b_1)}(\nu,\theta) = \int_{\frac{2\pi y_1}{2^{b_1}}-\pi-\theta}^{\frac{2\pi(y_1+1)}{2^{b_1}}-\pi-\theta}\Bigg[\tau\left(g_{y_2+1}^2,\phi,\nu\right)-  \tau\left(g_{y_2}^2,\phi,\nu\right)\Bigg]\;d\phi,
\end{equation}
where
\begin{equation}\label{eq:rho_func}
     \tau\left(r,\phi,\nu\right) = -\frac{e^{-r-\nu+2\sqrt{r\nu}\cos\phi}}{2\pi}-\frac{\sqrt{\nu}\cos\phi \;e^{-\nu\sin^2\phi}\left[1-2Q\left(\sqrt{2\nu}\cos\phi-\sqrt{2r}\right)\right]}{
2\sqrt{\pi}}.\end{equation}
\end{figure*}
\end{definition}

Mathematically, $W_{y_1,y_2}^{(b_1)}(\nu,\theta)$ is the probability that a complex Gaussian random variable with mean $\sqrt{\nu}\angle \theta$ and unit variance is inside the region bounded by the polar curves
\[r = g_{y_2+1}\;\mathrm{ for }\;\phi\in\left[\frac{2\pi y_1}{2^{b_1}},\frac{2\pi (y_1+1)}{2^{b_1}}\right]\]
and
\[r = g_{y_2}\;\mathrm{ for }\;\phi\in\left[\frac{2\pi y_1}{2^{b_1}},\frac{2\pi (y_1+1)}{2^{b_1}}\right].\]

In the next section, we will establish the operational meaning of $W_{y_1,y_2}^{(b_1)}(\nu,\theta)$. Essentially, when $Z\sim~ \mathcal{CN}(0,1)$ and $\mathcal{G}_{b_2}$ contains the magnitude quantizer thresholds, equation (\ref{eq:phase_quantization_prob})  describes the conditional probability mass function (PMF) of the $(b_1,b_2)$-bit polar-quantized channel outputs $(Y_1,Y_2)$ given $X = \sqrt{\nu}e^{j\theta}$. The integral for $\phi$ cannot be evaluated in closed-form. However, we can still identify the general structure of the optimal input using (\ref{eq:phase_quantization_prob}) and numerically compute the exact value of the capacity.

We now formally state the main result of this paper.

\begin{theorem}\label{theorem:AWGN_case} Under an average power constraint $P$ and nonzero phase quantization bits (i.e. $b_1 > 0$), the capacity of a complex Gaussian channel with $(b_1,b_2)$-bit polar quantizer at the output and with fixed channel gain $g_{\mathrm{LoS}}$ can be achieved by one of the following input structures:
\begin{itemize}
    \item \textbf{Constellation A:} A union of $L$ $2^{b_1}$-phase shift keying (PSK) constellations, where $L \leq 2^{b_2}$ and the $l$-th PSK constellation is given by the PMF
\begin{align}\label{eq:PMF_psk_ring}
    f^{(l)}_X(x) =& \Bigg\{\frac{\beta_l}{2^{b_1}}\;\;\bigg|\;\;x = \sqrt{\rho_l}e^{j\frac{2\pi k}{2^{b_1}}-\angle g_{\mathrm{LoS}}}, \nonumber\\
    &\;\forall k \in\{0,\cdots,2^{b_1}-1\}\Bigg\}
\end{align}
for some $\rho_l$ and $\beta_l$. Moreover, $\{\rho_l\}_{l=1}^{l=L}$ and $\{\beta_l\}_{l=1}^{l=L}$ should satisfy $\sum_{l=1}^{L}\beta_l = 1$ and the average power constraint $\sum_{l=1}^{L}\rho_l\beta_l = P$.

\item \textbf{Constellation B:}  A union of $L$ $2^{b_1}$-phase shift keying (PSK) constellations and a mass point at the origin ($\rho_0 = 0$) with probability $\beta_0$, where $L \leq 2^{b_2}-1$. The $l$-th PSK constellation is given by the PMF in (\ref{eq:PMF_psk_ring}) for some $\rho_l$ and $\beta_l$. Moreover, $\{\rho_l\}_{l=1}^{l=L}$ and $\{\beta_l\}_{l=1}^{l=L}$ should satisfy $\sum_{l=0}^{L}\beta_l = 1$ and the average power constraint $\sum_{l=1}^{L}\rho_l\beta_l = P$.
\end{itemize}
Consequently, the channel capacity can be expressed as \eqref{eq:cap_thm1}\footnote{All $\log(\cdot)$ functions in this paper are in base 2 unless stated otherwise.}. The function $Q_1(\cdot,\cdot)$ is the first-order Marcum-Q function. We set $\beta_0 = 0$ if Constellation A is optimal and $\beta_0 > 0$ otherwise.
\begingroup
\allowdisplaybreaks
\begin{figure*}[b]
\hrulefill
\begin{align}\label{eq:cap_thm1}
    C =&\;  b_1 - \sum_{y_2 = 0}^{2^{b_2}-1}\left[\sum_{l=0}^{L}\beta_lV_{y_2}\left(|g_{\mathrm{LoS}}|^2\rho_l,\frac{\sigma}{\sqrt{2}}\right)\right]\log\left[\sum_{l=0}^{L}\beta_lV_{y_2}\left(|g_{\mathrm{LoS}}|^2\rho_l,\frac{\sigma}{\sqrt{2}}\right)\right]\nonumber\\
    &+\sum_{y_1 = 0}^{2^{b_1}-1}\;\;\sum_{y_2 = 0}^{2^{b_2}-1}\sum_{l=0}^{L}\beta_lW_{y_1,y_2}^{(b_1)}\left(\frac{|g_{\mathrm{LoS}}|^2\rho_l}{\sigma^2},\frac{\pi}{2^{b_1}}\right)\log W_{y_1,y_2}^{(b_1)}\left(\frac{|g_{\mathrm{LoS}}|^2\rho_l}{\sigma^2},\frac{\pi}{2^{b_1}}\right),
\end{align}
where $V_{y_2}(\cdot,\cdot)$ is defined as
\begin{align}\label{eq:V_y2}
    V_{y_2}\left(t,\frac{\sigma}{\sqrt{2}}\right) = Q_{1}\left(\frac{\sqrt{t}}{\sigma/\sqrt{2}},\frac{q_{y_2}}{\sigma/\sqrt{2}}\right) - Q_{1}\left(\frac{\sqrt{t}}{\sigma/\sqrt{2}},\frac{q_{y_2+1}}{\sigma/\sqrt{2}}\right),
\end{align}
\end{figure*}
\endgroup
\end{theorem}

Constellations A and B of Theorem \ref{theorem:AWGN_case} correspond to APSK and on-off APSK modulation, respectively. To this end, we use the notation $(2^{b_1},L)$-APSK to refer to the specific structure of constellation A and on-off $(2^{b_1},L)$-APSK to refer to constellation B. For the special case of $L = 1$, we simply use $2^{b_1}$-PSK (on-off $2^{b_1}$-PSK) for constellation A (constellation B). It is worth noting that although the general structure of the optimal input is known, the evaluation of the capacity is still nontrivial and requires numerical computation of the optimal set of magnitude values and associated probability masses. Nonetheless, the dimension of the capacity maximization problem does not increase with the number of phase quantization bits because of the established properties of the optimal input.

Theorem \ref{theorem:AWGN_case} can be easily extended to the Gaussian MISO channel with polar quantization at the output. Suppose the transmitter has $N_{\mathrm{t}}$ antennas and has an average power constraint $P$. The channel gain from the $i$-th transmit antenna to the receiver is denoted as $g_{i}$. Consequently, the received signal can be written as
\begin{align}
    (Y_1,Y_2) = \mathcal{Q}_{b_1,b_2}^{\mathrm{PLR}}\left(\mathbf{g}^H\mathbf{X} + Z\right),
\end{align}
where $\mathbf{g}\in\mathbb{C}^{N_{\mathrm{t}}\times 1}$ is the channel vector containing $g_i$'s and $\mathbf{X}\in\mathbb{C}^{N_{\mathrm{t}}\times 1}$ is the signal sent by the transmitter. The following corollary establishes the capacity of the polar-quantized Gaussian MISO channel.
\begin{corollary}\label{corollary:MISO} Under an average power constraint $P$, the capacity of $(b_1,b_2)$-bit polar-quantized Gaussian MISO channel is given by \eqref{eq:cap_cor1}.
\begingroup
\allowdisplaybreaks
\begin{figure*}[b]
\hrulefill
\begin{align}\label{eq:cap_cor1}
    C_{\mathrm{MISO}} =&\;  b_1 - \sum_{y_2 = 0}^{2^{b_2}-1}\left[\sum_{l=0}^{L}\beta_lV_{y_2}\left(||\mathbf{g}||^2\rho_l,\frac{\sigma}{\sqrt{2}}\right)\right]\log\left[\sum_{l=0}^{L}\beta_lV_{y_2}\left(||\mathbf{g}||^2\rho_l,\frac{\sigma}{\sqrt{2}}\right)\right]\nonumber\\
    &+\sum_{y_1 = 0}^{2^{b_1}-1}\;\;\sum_{y_2 = 0}^{2^{b_2}-1}\sum_{l=0}^{L}\beta_lW_{y_1,y_2}^{(b_1)}\left(\frac{||\mathbf{g}||^2\rho_l}{\sigma^2},\frac{\pi}{2^{b_1}}\right)\log W_{y_1,y_2}^{(b_1)}\left(\frac{||\mathbf{g}||^2\rho_l}{\sigma^2},\frac{\pi}{2^{b_1}}\right),
\end{align}
\end{figure*}
\endgroup
\end{corollary}
\begin{proof}
The proof is similar to the proof of \cite[Proposition 1]{Mo:2014} for the 1-bit I/Q MISO channel. By Cauchy-Schwarz inequality, setting $\mathbf{X} = \frac{\mathbf{g}}{||\mathbf{g}||}S$, where $S$ is the information-bearing signal, maximizes the mutual information $I(S;Y_1,Y_2)$. Effectively, the polar-quantized MISO channel is transformed to an equivalent polar-quantized SISO channel with channel gain $||\mathbf{g}||$. That is,
\begin{align*}
    (Y_1,Y_2) = \mathcal{Q}_{b_1,b_2}^{\mathrm{PLR}}\left(||\mathbf{g}||S + Z\right).
\end{align*}
The corollary then follows by letting $S$ be the capacity-achieving input distribution in Theorem \ref{theorem:AWGN_case}.
\end{proof}

The proof of Theorem \ref{theorem:AWGN_case} is presented in Section \ref{section-thm1_proof}. Theorem \ref{theorem:AWGN_case} is then used in Section \ref{section-numerical_analysis} to numerically evaluate the capacity of AWGN channel with $(b_1,1)$-bit polar quantizer at the output and investigate the capacity-achieving input distribution in different SNR regimes. 

\section{Deriving the Capacity-Achieving Input: Proof of Theorem \ref{theorem:AWGN_case}}\label{section-thm1_proof}

The relationship between the $(b_1,b_2)$-bit polar quantizer outputs and the channel input can be written as
\begin{align}
    Y_1 =& \mathcal{Q}_{b_1}^{\mathrm{PH}}(Z) = \mathcal{Q}_{b_1}^{\mathrm{PH}}(g_{\text{LoS}}X+N)\\
    Y_2 =& \mathcal{Q}_{b_2}^{\mathrm{MG}}(Z) = \mathcal{Q}_{b_2}^{\mathrm{MG}}(g_{\text{LoS}}X+N)
\end{align}

Suppose we define $U = g_{\text{LoS}}X$ with the polar form $U = \sqrt{A}e^{j\Theta}$. Without loss of generality, we can simply find the capacity-achieving distribution for $U$ and apply the transformation $X = U/ g_{\text{LoS}}$.  The conditional PMF $p_{Y_1,Y_2|U}(y_1,y_2|u)$ (or $p_{Y_1,Y_2|A,\Theta}(y_1,y_2|\alpha,\theta)$) is
\begingroup
\allowdisplaybreaks
\begin{align}
    &p_{Y_1,Y_2|A,\Theta}(y_1,y_2|\alpha,\theta)\nonumber \\
    &= \int_{\mathcal{R}_{y_1}^{\text{PH}}}\int_{\mathcal{R}_{y_2}^{\text{MG}}}p_{R,\Phi|U}(r,\phi|u = \sqrt{\alpha}e^{j\theta})\;drd\phi\nonumber\\
    &= \int_{\mathcal{R}_{y_1}^{\text{PH}}}\int_{\mathcal{R}_{y_2}^{\text{MG}}}\frac{1}{2\pi\sigma^2}\exp\left(-\frac{|\sqrt{r}e^{j\phi}- \sqrt{\alpha}e^{j\theta}|^2}{\sigma^2}\right)\;drd\phi\nonumber\\
    &= \int_{\mathcal{R}_{y_1}^{\text{PH}}-\theta}\int_{\mathcal{R}_{y_2}^{\text{MG}}}\frac{1}{2\pi\sigma^2}\exp\left(-\frac{r+\alpha - 2\sqrt{r\alpha}\cos\phi}{\sigma^2}\right)\;drd\phi\nonumber\\
    &=\int_{\mathcal{R}_{y_1}^{\text{PH}}-\theta}\Bigg[\tau\left(\frac{q_{y_2+1}^2}{\sigma^2},\phi,\nu\right)-  \tau\left(\frac{q_{y_2}^2}{\sigma^2},\phi,\nu\right)\Bigg]\;d\phi\nonumber\\
    &= W_{y_1,y_2}^{(b_1)}\left(\frac{\alpha}{\sigma^2},\theta\right),
\end{align}
\endgroup
where the third line is obtained by rotating the whole problem by $\theta$ and expanding the expression in the exponent. The fourth and last lines are obtained from Definition \ref{definition:polar_quantization_func}, with the set $\mathcal{G}_{b_2}$ being $\mathcal{G}_{b_2} = \left\{\frac{q_{l}}{\sigma} \big| l\in\{0,\cdots,2^{b_2}\}\right\}$ (i.e. $\{g_l\}_{l = 0}^{l=2^{b_2}}$ in Definition \ref{definition:polar_quantization_func} are the magnitude thresholds $\{q_l\}_{l = 0}^{l=2^{b_2}}$ scaled by $\sigma^{-1}$). Now, consider a complex-valued distribution 
\begin{align*}
    F_U(u) = F_{A,\Theta}(\alpha,\theta) = F_A(\alpha)\cdot F_{\Theta|A}(\theta|\alpha),
\end{align*}
where $F_A(\alpha)$ and $F_{\Theta|A}(\theta|\alpha)$ are the amplitude distribution and phase distribution (conditioned on the amplitude) of the $U$, respectively. With slight abuse of notation, we use $F_U$ to refer to $F_U(u)$. For a given $F_U$, the joint PMF of $(Y_1,Y_2)$ is
\begin{equation}\label{eq:pmf_y}
        \begin{split}
             p(y_1,y_2;F_U) =& \int_{\mathbb{C}} W_{y_1,y_2}^{(b_1)}(u)\;dF_U\;\;\forall y_1,y_2,
        \end{split}
\end{equation}
where $W_{y_1,y_2}^{(b_1)}(u)$ is another way to write $W_{y_1,y_2}^{(b_1)}\left(\frac{\alpha}{\sigma^2},\theta\right)$ using the mapping $U = \sqrt{A}e^{j\Theta}$. These notations for $W_{y_1,y_2}^{(b_1)}(\cdots)$ will be used interchangeably. We also use the above notation for the joint PMF of $(Y_1,Y_2)$ to emphasize that it is induced by the choice of the distribution $F_U$. Given the above probability quantities, we can now express the mutual information between $U$ and $(Y_1,Y_2)$ as follows:
\begingroup
\allowdisplaybreaks
\begin{align}\label{eq:MI_U_Y}
        I(U;Y_1,Y_2) =& I(F_U) = H\left(Y_1,Y_2\right) - H\left(Y_1,Y_2|U\right),
\end{align}
where
\begin{align*}
    &H(Y_1,Y_2)\\
    &\qquad= -\int_{\mathbb{C}}\sum_{y_1=0}^{2^{b_1}-1}\sum_{y_2=0}^{2^{b_2}-1}W_{y_1,y_2}^{(b_1)}(u)\log p(y_1,y_2;F_U)\;dF_U
\end{align*}
and
\begin{align*}
    &H(Y_1,Y_2|U)\\
    &\qquad= -\int_{\mathbb{C}}\sum_{y_1=0}^{2^{b_1}-1}\sum_{y_2=0}^{2^{b_2}-1}W_{y_1,y_2}^{(b_1)}(u)\log W_{y_1,y_2}^{(b_1)}(u)\;dF_U.
\end{align*}
\endgroup
We introduced the notation $I(F_U)$ for the mutual information since this quantity is a result of choosing a specific distribution $F_U$. Thus, $I(F_U)$ and $I(U;Y_1,Y_2)$ can be used interchangeably. 

Let $P'=|g_{\text{LoS}}|^2P$. The capacity for a given power constraint is the supremum of mutual information between $U$ and $Y$ over the set of all distributions $F_U$ satisfying the power constraint $\mathbb{E}[|U|^2] \leq P'$. In other words,
\begin{equation}\label{eq:cap_def}
    C = \sup_{F_U \in \Omega} I(F_U) = I(F_U^*),
\end{equation}  
where $\Omega$ is the set of all input distributions which have average power less than or equal to $P'$ and $F_U^*\in\Omega$ is the capacity-achieving distribution. The mutual information $I(F_U)$ is concave with respect to $F_U$ \cite[Theorem 2.7.4]{Cover:2006_IT} and the power constraint ensures that $\Omega$ is convex and weakly compact with respect to weak* topology\footnote{This is the coarsest topology in which all linear functionals of $dF_U$ of the form $\int f(u)dF_U$, where $f(u)$ is a continuous function, are continuous.} \cite{Abou-Faycal:2001}. Moreover, because of the finite cardinality of the channel output, it is easy to verify that $I(F_U)$ is weak$^*$ continuous over $F_U$ and the proof follows closely to the method presented in \cite[Lemma 1]{Vu:2019}. Due to Theorem 2 of \cite[Section 5.10]{Luenberger:1968}, the existence of $F_U^*$ is guaranteed.



\subsection{Optimality of $\frac{2\pi}{2^{b_1}}$-symmetric distribution}

Given that an optimal distribution exists in the set $\Omega$, we now focus our attention on identifying the optimality conditions that an input distribution should satisfy. In this subsection, we focus on the underlying phase symmetry of $F_U^*$. We first present a lemma about $W_{y_1,y_2}^{(b_1)}(\nu,\theta)$.

\begin{lemma}\label{lemma:symmetry_Wy}
The function $W_{y_1,y_2}^{(b_1)}(\nu,\theta)$ has the following property:
\begin{align}
    W_{y_1,y_2}^{(b_1)}\left(\nu,\theta + \frac{2\pi k}{2^{b_1}}\right) = W_{y_1-k,y_2}^{(b_1)}\left(\nu,\theta\right), 
\end{align}
for any $k\in\mathbb{Z}$.
\end{lemma}
\begin{proof}
From the definition of $W_{y_1,y_2}^{(b_1)}(\nu,\theta)$, we have
\begingroup
\allowdisplaybreaks
\begin{align}
    &W_{y_1,y_2}^{(b_1)}\left(\nu,\theta + \frac{2\pi k}{2^{b_1}}\right) \\
    &= \int_{\frac{2\pi}{2^{b_1}}y_1 - \pi-\theta-\frac{2\pi k}{2^{b_1}}}^{\frac{2\pi}{2^{b_1}}(y_1+1) - \pi-\theta-\frac{2\pi k}{2^{b_1}}}\Big\{\tau\left(q_{y_2+1}^2,\phi,\nu\right)-  \tau\left(q_{y_2}^2,\phi,\nu\right)\Big\}\;d\phi\nonumber\\
    &=\int_{\frac{2\pi}{2^{b_1}}(y_1-k) - \pi-\theta}^{\frac{2\pi}{2^{b_1}}(y_1+1-k) - \pi-\theta}\Big\{\tau\left(q_{y_2+1}^2,\phi,\nu\right)-  \tau\left(q_{y_2}^2,\phi,\nu\right)\Big\}\;d\phi\nonumber\\
    &=W_{y_1-k,y_2}^{(b_1)}\left(\nu,\theta\right).
\end{align}
\endgroup
\end{proof}
Lemma \ref{lemma:symmetry_Wy} states that every shift of $\frac{2\pi k}{2^{b_1}}$ in the input distribution for some $k\in\mathbb{Z}$ is equivalent to a shift in the output of the phase quantizer component by $-k$. The first property of the optimal input distribution that we shall establish is its phase symmetry. Specifically, the capacity-achieving input should be a $\frac{2\pi}{2^{b_1}}$-symmetric distribution.
\begin{definition}\label{definition:2pi_K-symmetry}
Suppose $b_1 > 0$. A distribution $F_U$ is a $\frac{2\pi}{2^{b_1}}$-symmetric distribution if $F_U(u) \sim F_U(ue^{j\frac{2\pi k}{2^{b_1}}})$ for all $k\in\mathbb{Z}$. 
\end{definition}
To put it simply, a distribution that satisfies Definition \ref{definition:2pi_K-symmetry} will not change when any integer multiple rotation of $\frac{2\pi}{2^{b_1}}$ is applied to it. 
The first part of Proposition \ref{proposition:polar_symmetry_distribution} presents a transformation of any distribution to another distribution that satisfies Definition \ref{definition:2pi_K-symmetry}. We then show that this new distribution has the same conditional entropy as the original distribution yet attains higher output entropy. The proof is similar to that of \cite[Proposition 3]{bernardo2021TIT} in our previous work with an extra step of showing that $Y_1$ and $Y_2$ are independent when the input is $\frac{2\pi}{2^{b_1}}$-symmetric..

\begin{proposition}\label{proposition:polar_symmetry_distribution}
For any input distribution $F_U = F_A\cdot F_{\Theta|A}$, we define another distribution as
\begin{equation}
    F_U^{s} = \frac{1}{2^{b_1}}\sum_{i = 0}^{2^{b_1}-1}F_U(ue^{j\frac{2\pi i}{2^{b_1}}}),
\end{equation}
which is a $\frac{2\pi}{2^{b_1}}$-symmetric distribution. Then, $I(F_U^{s}) \geq I(F_U)$. Under this distribution, The output entropy $H(Y_1,Y_2)$ is maximized and is equal to $b_1$ + $H(Y_2)$ for some fixed distribution $F_A$.
\end{proposition}
\begin{proof}
See Appendix \ref{proof_symmetry}.
\end{proof}

Due to Proposition \ref{proposition:polar_symmetry_distribution}, the capacity can be expressed as
\begingroup
\allowdisplaybreaks
\begin{align}\label{eq:cap_sym}
    C 
    =& b_1 + \sup_{F_U\in\Omega_s} \Bigg\{-\sum_{y_2 = 0}^{2^{b_2}-1}p(y_2;F_A)\log p(y_2;F_A) \\
    &+ \int_{\mathbb{C}}\sum_{y_1=0}^{2^{b_1}-1}\sum_{y_2=0}^{2^{b_2}-1}W_{y_1,y_2}^{(b_1)}(u)\log W_{y_1,y_2}^{(b_1)}(u)\;dF_U\Bigg\}\nonumber\\
    =& b_1 + \sup_{F_U\in\Omega_s} \int_{\mathbb{C}}\sum_{y_1=0}^{2^{b_1}-1}\sum_{y_2=0}^{2^{b_2}-1}W_{y_1,y_2}^{(b_1)}(u)\log \frac{W_{y_1,y_2}^{(b_1)}(u)}{p(y_2;F_A)}\;dF_U,
\end{align}
\endgroup
where $p(y_2;F_A)$ is the marginal PMF of $Y_2$ induced by the choice of amplitude distribution $F_A$, and $\Omega_s$ is the set of all $\frac{2\pi}{2^{b_1}}$-symmetric distributions satisfying the average power constraint. The last line is due to the fact that
\begin{align*}
    p(y_2;F_A) =\int_{\mathbb{C}}\sum_{y_1=0}^{2^{b_1}-1}W_{y_1,y_2}^{(b_1)}(u)\;dF_U
\end{align*}
which follows from the arguments presented in Appendix \ref{proof_symmetry}. As such, both summation terms can be combined accordingly.

\subsection{Kuhn-Tucker Condition}

The use of Lagrange Multiplier Theorem in finding the optimal distribution of $U$ in this problem requires that the mutual information is weakly differentiable with respect to $F_U$. That is, for a given $F_U^0 \in \Omega_{s}$ and $\lambda \in [0,1]$, the quantity
\begin{equation}\label{eq:weak_derivative_I}
    I'_{F_U^0}(F_U) = \lim_{\lambda \rightarrow 0 }\frac{I\left((1-\lambda)F_U^0 + \lambda F_U\right) - I(F_U^0)}{\lambda}
\end{equation}
exists $\forall F_U \in \Omega_s$. Let $F_U^\lambda = (1-\lambda)F_U^0 + \lambda F_U$ and define the divergence function $d(u;F_U)$\footnote{\label{footnote:divergence} Alternatively, we can use the notation $d\left(\frac{\alpha}{\sigma^2},\theta;F_U\right)$. Both $d\left(\frac{\alpha}{\sigma^2},\theta;F_U\right)$ and $d\left(u;F_U\right)$ can be used interchangeably.} as
\begin{align}\label{eq:divergence_func}
    d(u;F_U) = \sum_{y_1=0}^{2^{b_1}-1}\sum_{y_2=0}^{2^{b_2}-1}W_{y_1,y_2}^{(b_1)}(u)\log \frac{W_{y_1,y_2}^{(b_1)}(u)}{p(y_2;F_A)}.
\end{align}
The weak derivative can be expressed as
\begingroup
\allowdisplaybreaks
\begin{align*}
    &I'_{F_U^0}(F_U) \\
    &\quad= \lim_{\lambda \rightarrow 0 } \frac{\int_{C}d(u;F_U^\lambda)\;dF_U^\lambda - \int_{C}d(u;F_U)\;dF_U^0}{\lambda}\\
    &\quad= \lim_{\lambda \rightarrow 0 } \frac{(1-\lambda)\int_{C}d(u;F_U^\lambda)\;dF_U^0}{\lambda}\\
    &\quad\quad+\frac{\lambda\int_{C}d(u;F_U^\lambda)\;dF_U- \int_{C}d(u;F_U)\;dF_U^0}{\lambda}\\
    &\quad= \int_{C}d(u;F_U^0)\;dF_U -\int_{C}d(u;F_U^0)\;dF_U^0 \\
    &\quad\quad+  \lim_{\lambda \rightarrow 0 }\frac{\int_{C} \sum_{y_1=0}^{2^{b_1}-1}\sum_{y_2=0}^{2^{b_2}-1}W_{y_1,y_2}^{(b_1)}(\alpha)\log \frac{p(y_2;F_A)}{p(y_2;F_A^\lambda)}\;dF_U^0}{\lambda}.
\end{align*}
\endgroup
Since $p(y_2;F_A^\lambda) =  (1-\lambda)p(y_2;F_A^0) + \lambda p(y_2;F_A)$, it can be shown by L'hopital's Rule that the last term vanishes as $\lambda \rightarrow 0$. Thus, we have
\begingroup
\allowdisplaybreaks
\begin{align*}
    I'_{F_U^0}(F_U) =& \int_{C}d(u;F_U^0)\;dF_U -\int_{C}d(u;F_U^0)\;dF_U^0
\end{align*}
\endgroup
which exists because both terms are finite. Combining the weakly differentiable property of $I(F_U)$ with the concavity of $I(F_U)$ and convexity and compactness of $\Omega_s$ implies the existence of a non-negative Lagrange multiplier $\mu$ such that
\begin{align*}
    C = \sup_{F_U\in \Omega_{s}} I(F_U) = \sup_{F_U\in\Omega_{s}^0} \;I(F_U) - \mu \phi(F_U),
\end{align*}
where $\phi(F_U) = \int |u|^2dF_U - P'$ and $\Omega_{s}^0$ is the set of all $\frac{2\pi}{2^b_1}$-symmetric distributions. It is easy to show that $\phi(F_U)$ is also weakly differentiable over $F_U$ (i.e. $\phi_{F_U^0}'(F_U) = \phi(F_U) - \phi(F_U^0)$) and so is $I(F_U)-\mu\phi(F_U)$. Moreover, since $\phi(F_U)$ is linear in $F_U$ and $I(F_U)$ is concave in $F_U$, then $I(F_U)-\mu\phi(F_U)$ is also concave in $F_U$. Thus, $F_U^* \in \Omega_s$ is optimal if for all $F_U$, we have
\begingroup
\allowdisplaybreaks
\begin{align*}
        I'_{F_U^*}(F_U) - \mu\phi'_{F_U^*}(F_U) \leq& 0\\
        b_1 + \int_{C}d(u;F_U^*)\;dF_U - \mu\int_{\mathbb{C}}|u|^2\;dF_U \leq& C - \mu P',
\end{align*} 
\endgroup
where we used (\ref{eq:cap_def}), (\ref{eq:cap_sym}), and the complementary slackness of the constraint (having $\int_{\mathbb{C}}|u|^2dF_U^*$ strictly less than $P'$ makes $\mu = 0$ and the expression still holds) in the last inequality. Finally, using the same contradiction argument in \cite[Theorem 4]{Abou-Faycal:2001}, noting that $|u| = \sqrt{\alpha}$, and after some algebraic manipulation, the KTC can be established as
\begin{align}\label{eq:KTC}
  C - b_1 + \mu(\alpha- P') - d(u;F_U^*) \geq 0,
\end{align}
and equality is achieved when $u = \sqrt{\alpha}e^{j\theta}$ is a mass point of $F_U^*$. The
KTC will be used to prove some properties of $F_U^*$ as well as
identify which mass points belong to $F_U^*$.

\subsection{Boundedness and Discreteness of the Optimal Distribution}

The boundedness of the optimal input is proven using the KTC. The key idea is to consider two scenarios of the Lagrange multiplier (i.e. $\mu = 0$ and $\mu > 0$) and show that in both cases, equality in (\ref{eq:KTC}) cannot be achieved if $\alpha \rightarrow \infty$.

\begin{lemma}\label{lemma:boundedness_input}
The optimal distribution $F_U^*$ has a bounded support.
\end{lemma}
\begin{proof}
See Appendix \ref{proof_boundedness}.
\end{proof}

We use this boundedness property in Proposition \ref{proposition:discreteness_input} to
show that $F_U$ is discrete and identify an upper bound on the number of mass points. The proof technique follows closely from the approach used by \cite[Section V-B]{Vu:2019} and \cite[Proposition 1]{Rahman2:2020} to prove that the optimal input of a noncoherent Rician channel with $K$-bit I/Q ADC and a zero-mean Gaussian mixture channel with 1-bit I/Q ADC should be discrete distributions with at most $2^{2K}$ mass points and at most 4 mass points, respectively. We use the fact that $I(F_U) - \mu\phi(F_U)$ is a linear functional
of the bounded $F_U$. Thus, Dubins' Theorem \cite{Dubins:1962} can be applied. 

\begin{proposition}\label{proposition:discreteness_input}
The optimal input $F_U^*$ has a discrete support set with at most $2^{b_1+b_2}$ mass points.
\end{proposition}

\begin{proof}
See Appendix \ref{proof_discreteness}.
\end{proof}

Both the discreteness and boundedness properties can be exploited by gradient-based \cite{Abou-Faycal:2001} and cutting-plane-based algorithms \cite{Huang:2005} to numerically search for the location of these mass points. While an upper bound of $2^{b_1+b_2}$ is established in Proposition \ref{proposition:discreteness_input}, we show in the next section that the complexity of numerical approaches to find these mass points does not need to scale with $b_1$ since the phase information of these mass points can be solved analytically.

\subsection{Angles of the Optimal Mass Points not located at the Origin}

We now examine the optimal angles of the mass points not located at the origin using the KTC. We first identify some symmetry properties of $W_{y_1,y_2}^{(b_1)}(\nu,\theta)$ for $\theta = 0$ and $\theta = \frac{\pi}{2^{b_1}}$.

\begin{lemma}\label{lemma:specialvalues_Wy}
    The function $W_{y_1,y_2}^{(b_1)}(\nu,\theta)$ has the following symmetry for $\theta = 0$ and $\theta = \frac{\pi}{2^{b_1}}$:
\begin{align}
    (i) && W_{2^{b_1-1}-y_1,y_2}^{(b_1)}\left(\nu,\frac{\pi}{2^{b_1}}\right) =& W_{2^{b_1-1}+y_1,y_2}^{(b_1)}\left(\nu,\frac{\pi}{2^{b_1}}\right)\\
    (ii) && W_{2^{b_1-1}-y_1,y_2}^{(b_1)}\left(\nu,0\right) =& W_{2^{b_1-1}-1+y_1,y_2}^{(b_1)}\left(\nu,0\right)
\end{align}
\end{lemma}

\begin{proof}
From the definition of $W_{y_1,y_2}^{(b_1)}(\nu,\theta)$, we have
\begingroup
\allowdisplaybreaks
\begin{align}
    &W_{2^{b_1-1}-y_1,y_2}^{(b_1)}\left(\nu,\frac{\pi}{2^{b_1}}\right) \nonumber\\
    &\quad=\int_{-\frac{2\pi y_1}{2^{b_1}} - \frac{\pi}{2^{b_1}}}^{-\frac{2\pi y_1}{2^{b_1}} + \frac{\pi}{2^{b_1}}}\Big\{\tau\left(q_{y_2+1}^2,\phi,\nu\right)-  \tau\left(q_{y_2}^2,\phi,\nu\right)\Big\}\;d\phi\nonumber\\
    &\quad=\int_{\frac{2\pi y_1}{2^{b_1}} - \frac{\pi}{2^{b_1}}}^{\frac{2\pi y_1}{2^{b_1}} + \frac{\pi}{2^{b_1}}}\Big\{\tau\left(q_{y_2+1}^2,\phi',\nu\right)-  \tau\left(q_{y_2}^2,\phi',\nu\right)\Big\}\;d\phi'\nonumber\\
    &\quad=W_{2^{b_1-1}+y_1,y_2}^{(b_1)}\left(\nu,\frac{\pi}{2^{b_1}}\right).
\end{align}
\endgroup
The third equality follows from letting $\phi' = -\phi$. This proves Lemma \ref{lemma:specialvalues_Wy}.i. Note that the second line follows from a change of variable $\phi' = -\phi$ and the even symmetry of the $\cos(\cdot)$ and $\sin^2(\cdot)$  terms of $\tau(r,\phi,\nu)$. Meanwhile, for Lemma \ref{lemma:specialvalues_Wy}.ii, we have
\begingroup
\allowdisplaybreaks
\begin{align}
   &W_{2^{b_1-1}-y_1,y_2}^{(b_1)}\left(\nu,0\right) \nonumber\\ &\qquad= \int_{-\frac{2\pi}{2^{b_1}}y_1}^{-\frac{2\pi}{2^{b_1}}y_1 + \frac{2\pi}{2^{b_1}}}\Big\{\tau\left(q_{y_2+1}^2,\phi,\nu\right)-  \tau\left(q_{y_2}^2,\phi,\nu\right)\Big\}\;d\phi\nonumber\\
   &\qquad= \int_{\frac{2\pi}{2^{b_1}}y_1- \frac{2\pi}{2^{b_1}}}^{\frac{2\pi}{2^{b_1}}y_1 }\Big\{\tau\left(q_{y_2+1}^2,\phi',\nu\right)-  \tau\left(q_{y_2}^2,\phi',\nu\right)\Big\}\;d\phi'\nonumber\\
   &\qquad= W_{2^{b_1-1}-1+y_1,y_2}^{(b_1)}\left(\nu,0\right)
\end{align}
\endgroup
which completes the proof.
\end{proof}

We assume that $\alpha > 0$ and we limit the search of $\theta$ in $\left[0,\frac{2\pi}{2^{b_1}}\right)$ (i.e. $\theta \in \mathcal{R}_{2^{b_1-1}}^{\mathrm{PH}}$) since if $\theta^*$ is an angle of the optimal mass point, so are $\left\{\theta^* +\frac{2\pi k}{2^{b_1}}\right\}_{k = 1}^{2^{b_1}-1}$. Suppose we let $\mathcal{L}(\alpha,\theta,\mu)$ be the LHS of (\ref{eq:KTC}). That is,
\begin{align}\label{eq:Lagrangian}
    \mathcal{L}(\alpha,\theta,\mu) =&  C - b_1 + \mu(\alpha- P') \nonumber\\
    &- \sum_{y_1=0}^{2^{b_1}-1}\sum_{y_2=0}^{2^{b_2}-1}W_{y_1,y_2}^{(b_1)}\left(\frac{\alpha}{\sigma^2},\theta\right)\log \frac{W_{y_1,y_2}^{(b_1)}\left(\frac{\alpha}{\sigma^2},\theta\right)}{p(y_2;F_A^*)}.
\end{align}
The necessary (but not sufficient) conditions for $u^* = \sqrt{\alpha^*}e^{j\theta^*}$ to be a minimizer of $\mathcal{L}(\alpha,\theta,\mu)$ are the following:
\begin{align}
     \nabla_{\theta}\mathcal{L}(\alpha^*,\theta^*,\mu) =& 0\qquad\text{ and }\label{eq:stationary_phase}\\
     \qquad\nabla_{\alpha}\mathcal{L}(\alpha^*,\theta^*,\mu) =& 0.\label{eq:stationary_magnitude}
\end{align}
We use the stationary condition (\ref{eq:stationary_phase}) to find the angles of the optimal mass points. The proof technique is based on \cite[Appendix I]{bernardo2021TIT}.

\begin{proposition}\label{proposition:optimal_angles}
An optimal mass point not located at the origin should have an angle contained in the set
\begin{align}
    \mathbf{\Theta}^* = \left\{\frac{2\pi (k + 0.5)}{2^{b_1}}\right\}_{k =0}^{2^{b_1}-1}
\end{align}
In other words, the angles of the optimal  mass points should coincide with the angle bisector of the phase quantization regions.
\end{proposition}
 \begin{proof}
 See Appendix \ref{proof_optimal_angles}.
 \end{proof}

We now combine all the propositions to establish the structure of the optimal input. First, note that the optimal input is discrete and has at most $2^{b_1+b_2}$ mass points (Proposition \ref{proposition:discreteness_input}). To satisfy $\frac{2\pi}{2^{b_1}}$-symmetry (Proposition \ref{proposition:polar_symmetry_distribution}), these $2^{b_1+b_2}$ mass points should be distributed evenly to $2^{b_1}$ phase quantization regions $\mathcal{R}_{y_1}^{\mathrm{PH}}$ and so each $\mathcal{R}_{y_1}^{\mathrm{PH}}$ should have at most $2^{b_2}$ mass points. Finally, Proposition \ref{proposition:optimal_angles} makes sure that the optimal mass points at a phase quantization region are aligned at the middle of the phase quantization region. This gives an APSK structure (i.e. Constellation A) in Theorem \ref{theorem:AWGN_case}. By \cite[Theorem 2]{Agrell:2015}, the capacity of the channel is a non-decreasing function of $P$ and so, without loss of generality, we can simply consider APSK distributions that satisfy the average power constraint with equality. The same argument works for the case when an optimal mass point is located at the origin to get an on-off APSK (i.e. Constellation B). If one mass point is located at the origin, then at most $2^{b_2}-1$ can be placed in a phase quantization region to satisfy the $\frac{2\pi}{2^{b_1}}$-symmetry condition. Placing more will violate the symmetry. Using these input structures to evaluate (\ref{eq:cap_sym}) gives the capacity expression in (\ref{eq:cap_thm1}).

\section{Numerical Analysis of ($b_1$,1)-bit Polar-quantized AWGN Channel}\label{section-numerical_analysis}

In this section, we consider a simple case of AWGN channel with $(b_1,1)$-bit polar quantizer at the output and then use the established results in Theorem \ref{theorem:AWGN_case} to numerically compute the capacity.

\begin{figure*}[t]
    \centering
    \subfloat[]{
    \includegraphics[scale = .5625]{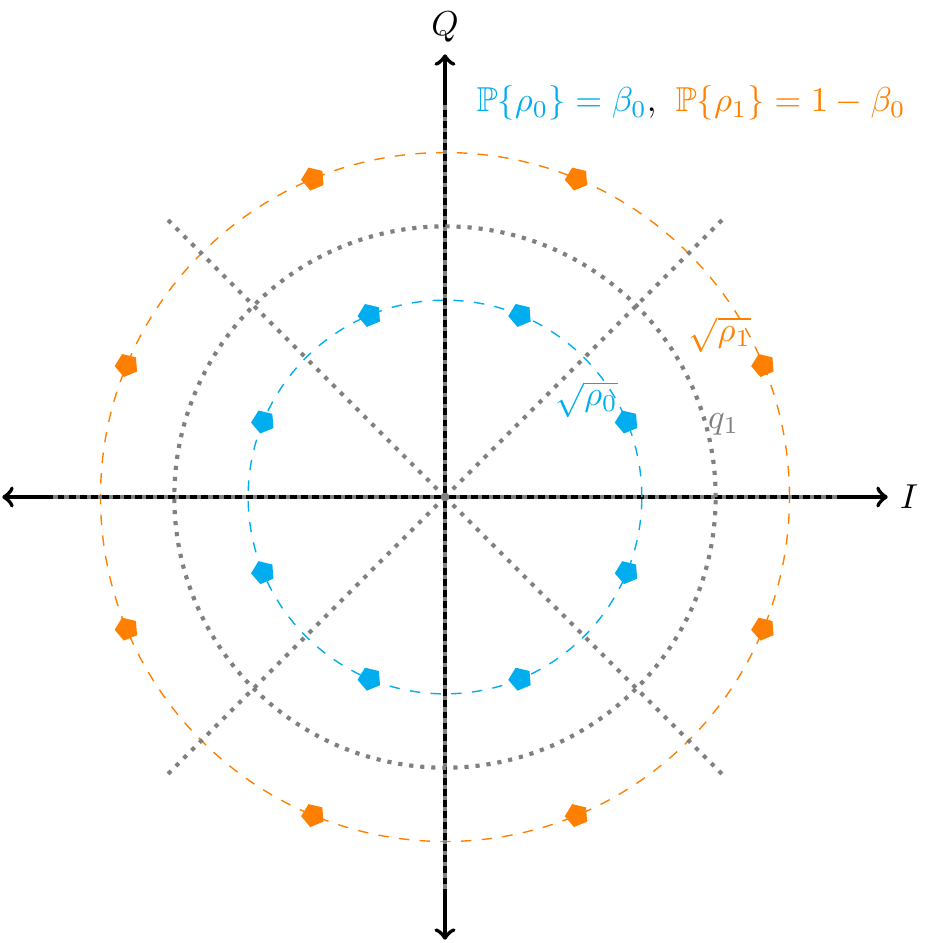}
    }
    \subfloat[]{
    \includegraphics[scale = .5625]{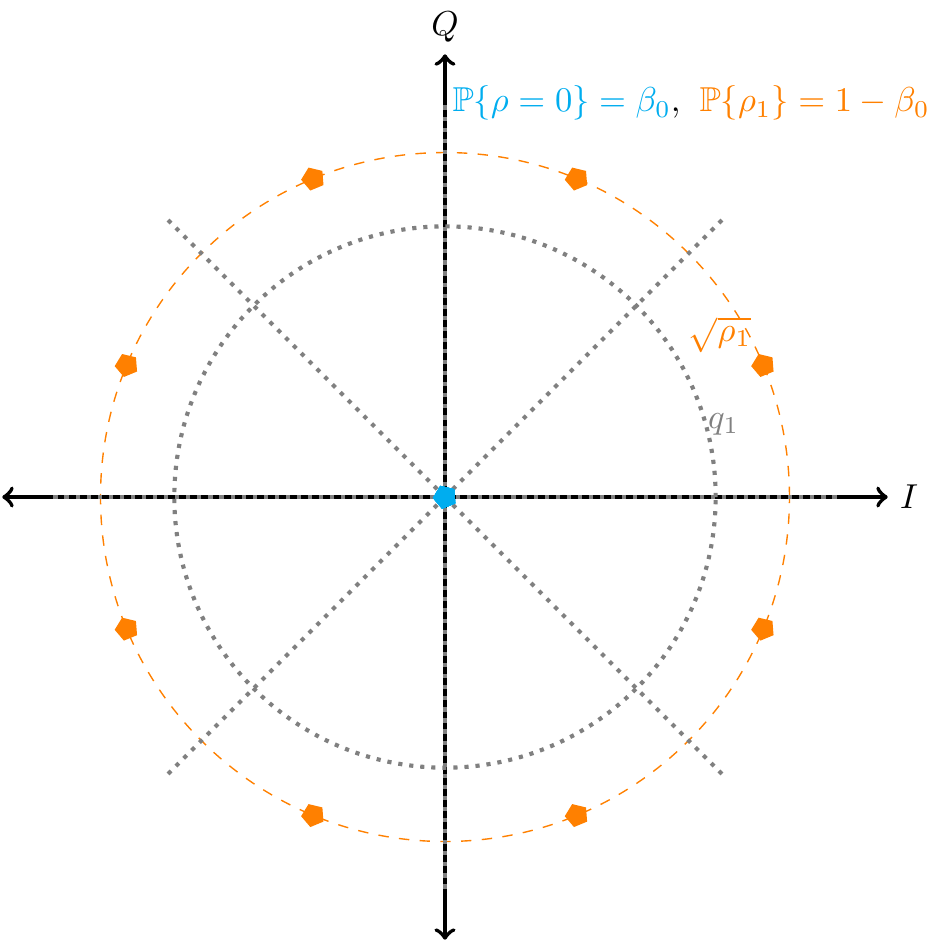}
    }
    \subfloat[]{
    \includegraphics[scale = .5625]{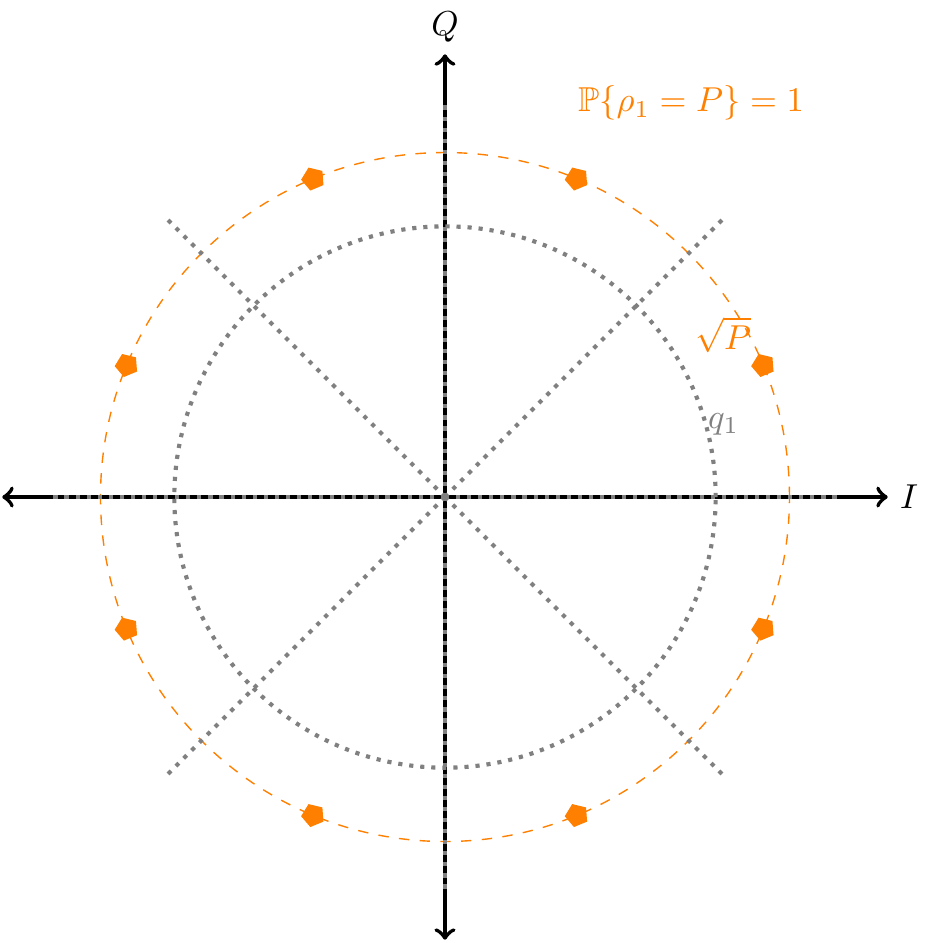}
    }
    \caption{Numerical Setup for the $(3,1)$-bit polar quantizer and (a) $(8,2)$-APSK. The gray lines and gray circle correspond to the phase quantization and magnitude quantization, respectively. Special cases are the (b) on-off 8-PSK and (c) 8-PSK.}
    \label{fig:apsk_polar_quant}
\end{figure*}

\subsection{Experiment Setup}
Without loss of generality, we assume $g_{\mathrm{LoS}} = 1$ and $P = 1$ so the SNR is varied by changing the noise variance of the additive noise. By Theorem \ref{theorem:AWGN_case}, the capacity-achieving input is an APSK with at most 2 amplitude levels and exactly $2^{b_1}$ phase values. We denote the ``lower" and ``upper" amplitude levels as $\sqrt{\rho_0}$ and $\sqrt{\rho_1} = \sqrt{\frac{P-\beta_0\rho_0}{\beta_1}}$, respectively, and their corresponding probabilities as $\beta_0$ and $\beta_1 = 1-\beta_0$. Furthermore, we also optimize the radial threshold, $q_1$, of the single-bit magnitude quantizer together with the channel input. To be more precise, we focus on optimization problem \eqref{eq:cap_numeric},
\begingroup
\allowdisplaybreaks
\begin{figure*}[b]
\hrulefill
\begin{align}\label{eq:cap_numeric}
    C =&\; \underset{\rho_0,\beta_0,q_1}{\max} \Bigg\{b_1 - \mathcal{H}_{b}\left\{\beta_0V_{y_2}\left(\rho_0,\frac{\sigma}{\sqrt{2}}\right)+(1-\beta_0)V_{y_2}\left(\frac{P-\beta_0\rho_0}{1-\beta_0},\frac{\sigma}{\sqrt{2}}\right)\right\}\nonumber\\
    &\qquad+\sum_{y_1 = 0}^{2^{b_1}-1}\sum_{y_2 = 0}^{1}\Bigg[\beta_0W_{y_1,y_2}^{(b_1)}\left(\frac{\rho_0}{\sigma^2},\frac{\pi}{2^{b_1}}\right)\log W_{y_1,y_2}^{(b_1)}\left(\frac{\rho_0}{\sigma^2},\frac{\pi}{2^{b_1}}\right)\nonumber\\
    &\qquad+(1-\beta_0)W_{y_1,y_2}^{(b_1)}\left(\frac{(P-\beta_0\rho_0)}{\sigma^2(1-\beta_0)},\frac{\pi}{2^{b_1}}\right)\log W_{y_1,y_2}^{(b_1)}\left(\frac{(P-\beta_0\rho_0)}{\sigma^2(1-\beta_0)},\frac{\pi}{2^{b_1}}\right)\Bigg]\Bigg\},
\end{align}
\end{figure*}
\endgroup
where $\mathcal{H}_{b}\{\cdot\}$ is the binary entropy function. An illustration of the numerical setup of a $(3,1)$-bit polar-quantized channel is depicted in Figure \ref{fig:apsk_polar_quant}. The setup can be readily extended to other $b_1$ by changing the number of phase quantization regions and APSK phase values. When $\rho_0 = 0$ or when $\beta_0 = 0$, the constellation collapses to an on-off PSK or PSK, respectively (see Figures \ref{fig:apsk_polar_quant}b and \ref{fig:apsk_polar_quant}c). 

Equation (\ref{eq:cap_numeric}) jointly optimizes the quantizer and input distribution. This problem, however, is known to be computationally intractable due to its nonconvex structure \cite{Kurkoski:2012}. We use an alternate iterative optimization procedure to identify the capacity-achieving input distribution (parametrized by $\rho_0$ and $\beta_0$) and the optimal quantizer (parametrized by $q_1$). More precisely, we specify an initial value of $q_1$ (say $q'_1$) then perform iteration as follows:
\begin{enumerate}
    \item  For a fixed $q_1'$, find the parameters $\rho_0$ and $\beta_0$ that describes the capacity-achieving input. Call this $(\rho'_0,\beta'_0)$.
    \item Using $(\rho'_0,\beta'_0)$ in step 1, find the optimal quantizer $q_1'$.
    \item Repeat the first two steps until the capacity gain is less than some threshold $\epsilon$.
\end{enumerate}
We use the gradient-based $\mathrm{fmincon(\cdot)}$ function of MATLAB to solve each optimization problem. The above scheme is not guaranteed to converge to the global optimal solution so we use multiple intializations of $q_1$ to improve the chance that the algorithm will converge to the best solution. While we found that gradient-based methods work well in our setup due to the small number of parameters we need to optimize, we note that such approach may be unstable in the case where the input has a lot of amplitude levels; especially when some of the amplitude levels have very low probability values. The performance of other existing approaches (e.g. Blahut-Arimoto \cite{Blahut:1972}, Cutting-plane-based methods \cite{Huang:2005}) in this setting can be further investigated but this is beyond the scope of the current work.

\subsection{Numerical Results and Discussion}

\begin{figure}[t]
    \centering
    \includegraphics[scale = .65]{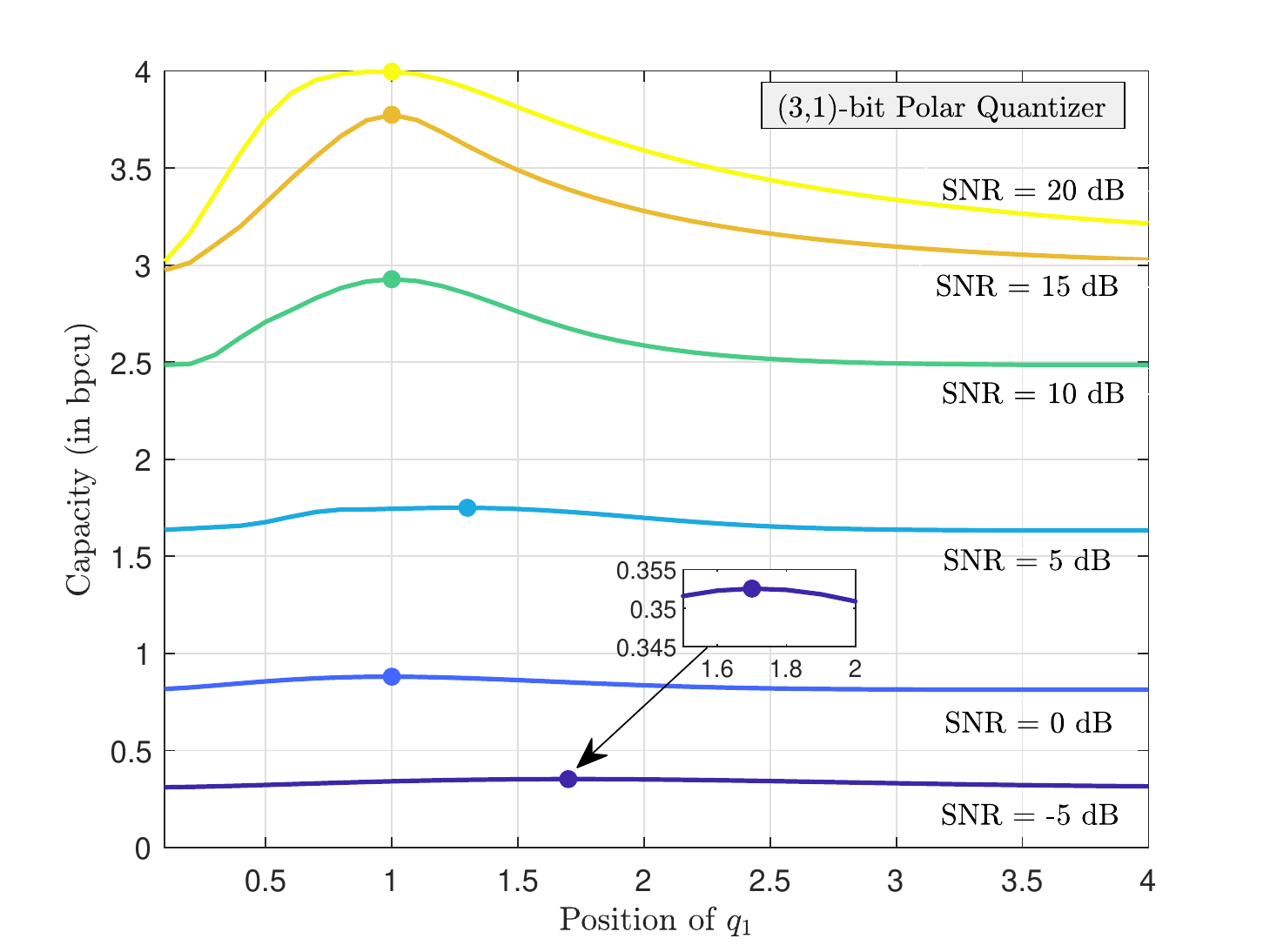}
    \caption{Capacity vs. $q_1$ of $(3,1)$-bit polar-quantized channels. Circle markers are the maximum point of the plots}
    \label{fig:cap_vs_pos}
\end{figure}

We first look at the variation of the channel capacity as a function of the quantizer. Figure \ref{fig:cap_vs_pos} shows the capacity of a $(3,1)$-bit polar-quantized channel as a function of $q_1$ for different SNR values. It can be observed that there is an optimal choice of $q_1$ that maximizes the capacity for any SNR. Moreover, the variation in the channel capacity is small in the low SNR regime but the variation becomes more pronounced as SNR is increased. As such, the quantizer choice becomes more crucial in the high SNR regime.

\begin{figure}[t]
    \centering
    \includegraphics[scale = .65]{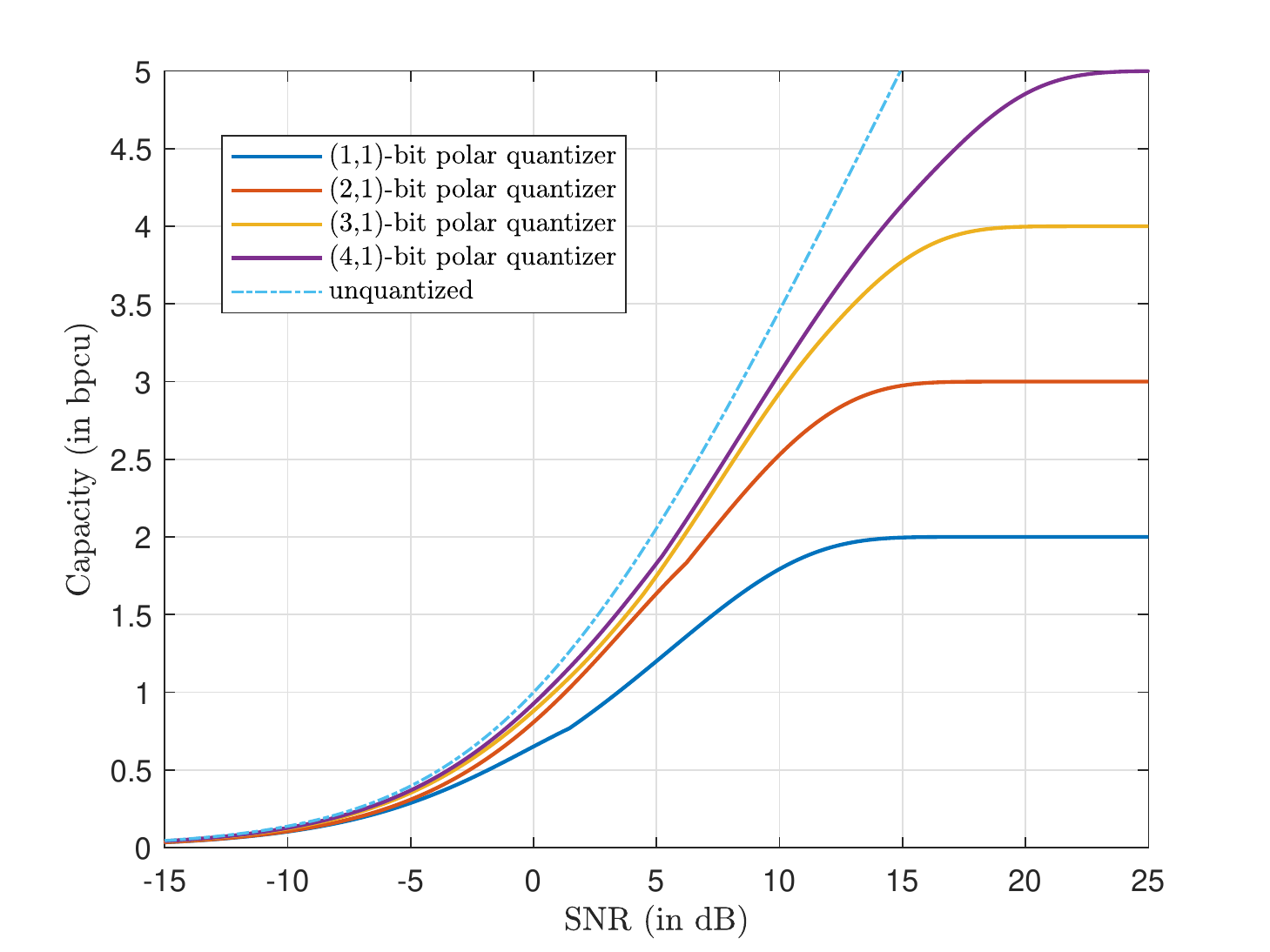}
    \caption{Capacity vs. SNR of $(b_1,1)$-bit polar-quantized channels}
    \label{fig:cap_vs_snr}
\end{figure}

We compare the capacity results of different precisions of $(b_1,1)$-bit polar-quantized AWGN channels in Figure \ref{fig:cap_vs_snr}. The capacity of the unquantized complex-valued AWGN channel is also superimposed in Figure \ref{fig:cap_vs_snr} to get an idea of how large the capacity loss is by using such quantization strategy. We observe that in the low SNR regime, the reduction in capacity is small. For example, at SNR = 0 dB, a $(2,1)$-bit polar-quantized channel already achieves 80.7\% of the unquantized AWGN capacity, while a $(3,1)$-bit polar-quantized channel gets around 88\% of the unquantized AWGN capacity. In the high SNR regime, the capacity of the polar-quantized channel is capped at $b_1 + b_2$ bits per channel use, which is the maximum value of the output entropy.

\begin{figure*}[t]
  \centering
  \hspace*{-0.75cm}
  \subfloat[]{
\includegraphics[width =.56\textwidth]{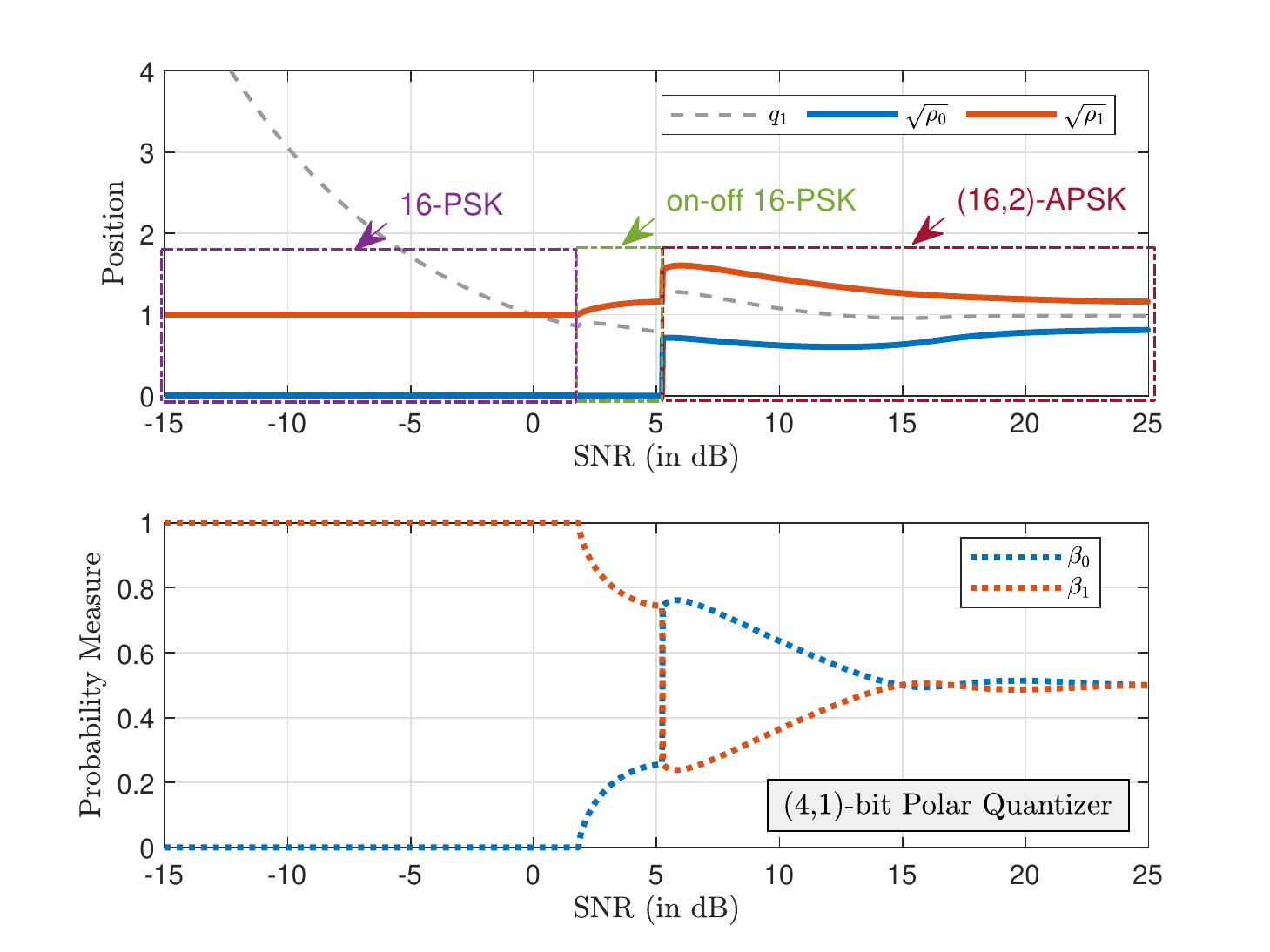}
\label{fig:optimal_input_4_1}
}
\hspace*{-1cm}%
 \subfloat[]{
\includegraphics[width =0.56\textwidth]{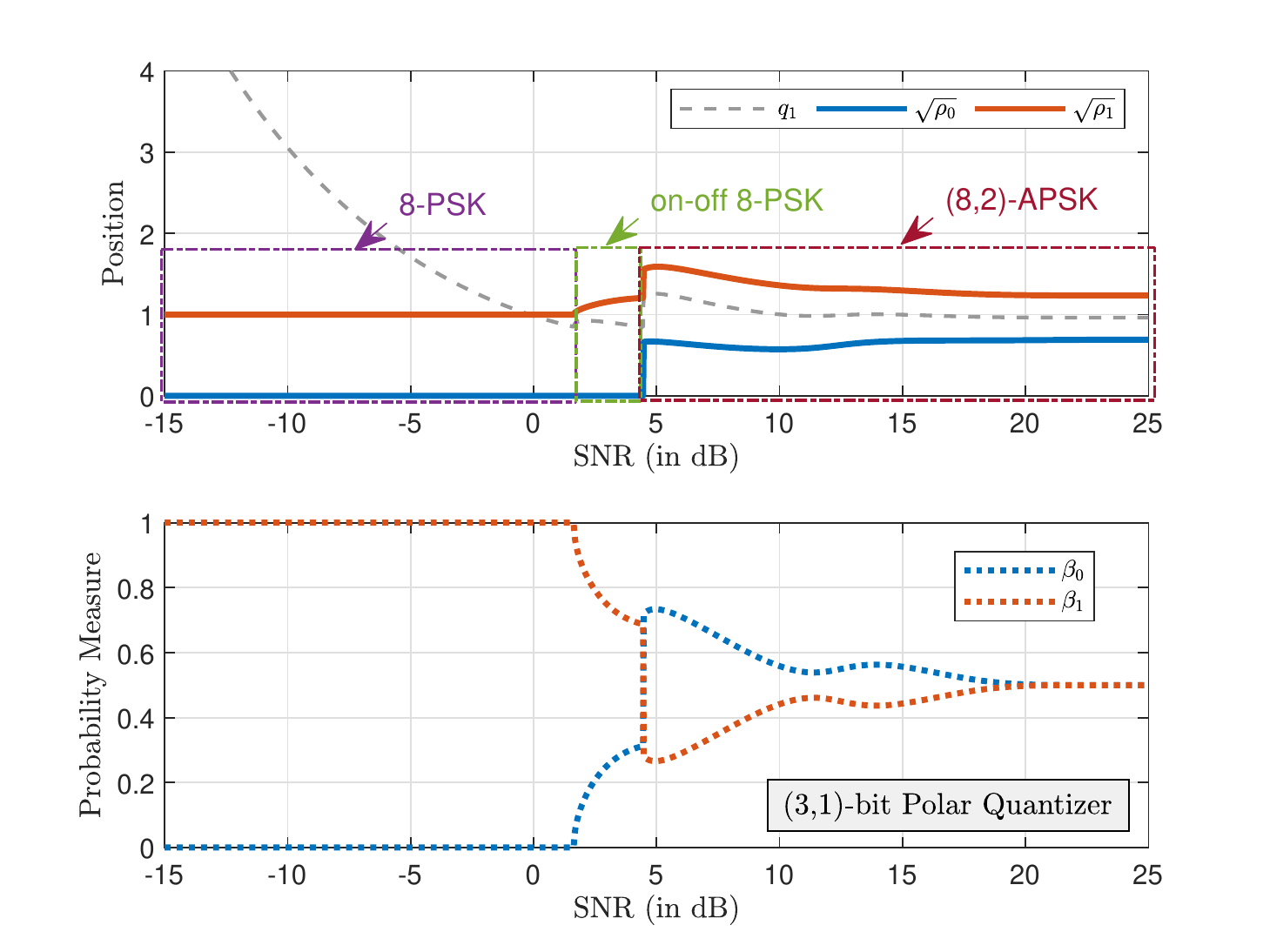}
    \label{fig:optimal_input_3_1}
} 
\\
\hspace*{-0.75cm}
  \subfloat[]{
\includegraphics[width =.56\textwidth]{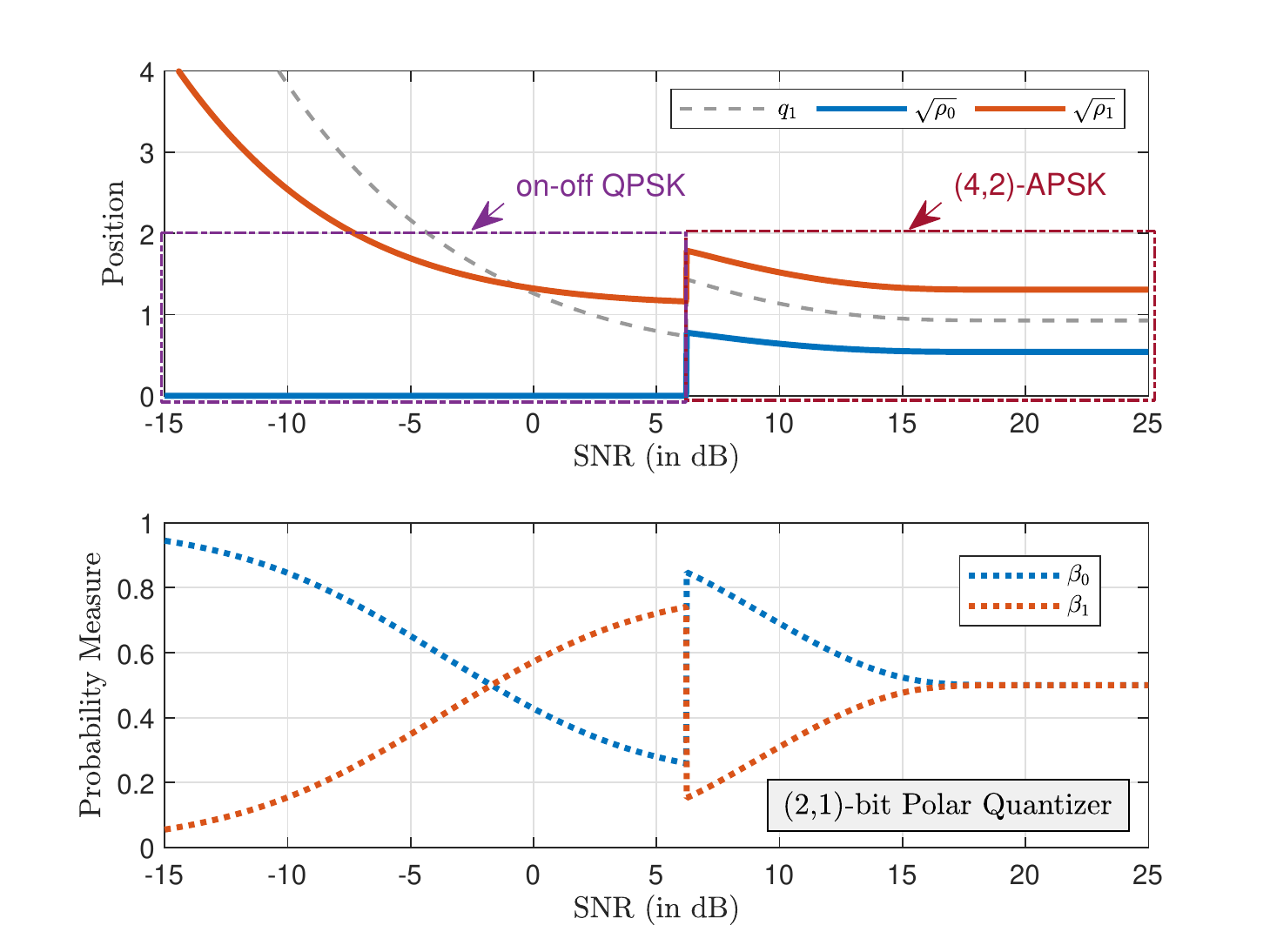}
\label{fig:optimal_input_2_1}
}
\hspace*{-1cm}%
 \subfloat[]{
\includegraphics[width =0.56\textwidth]{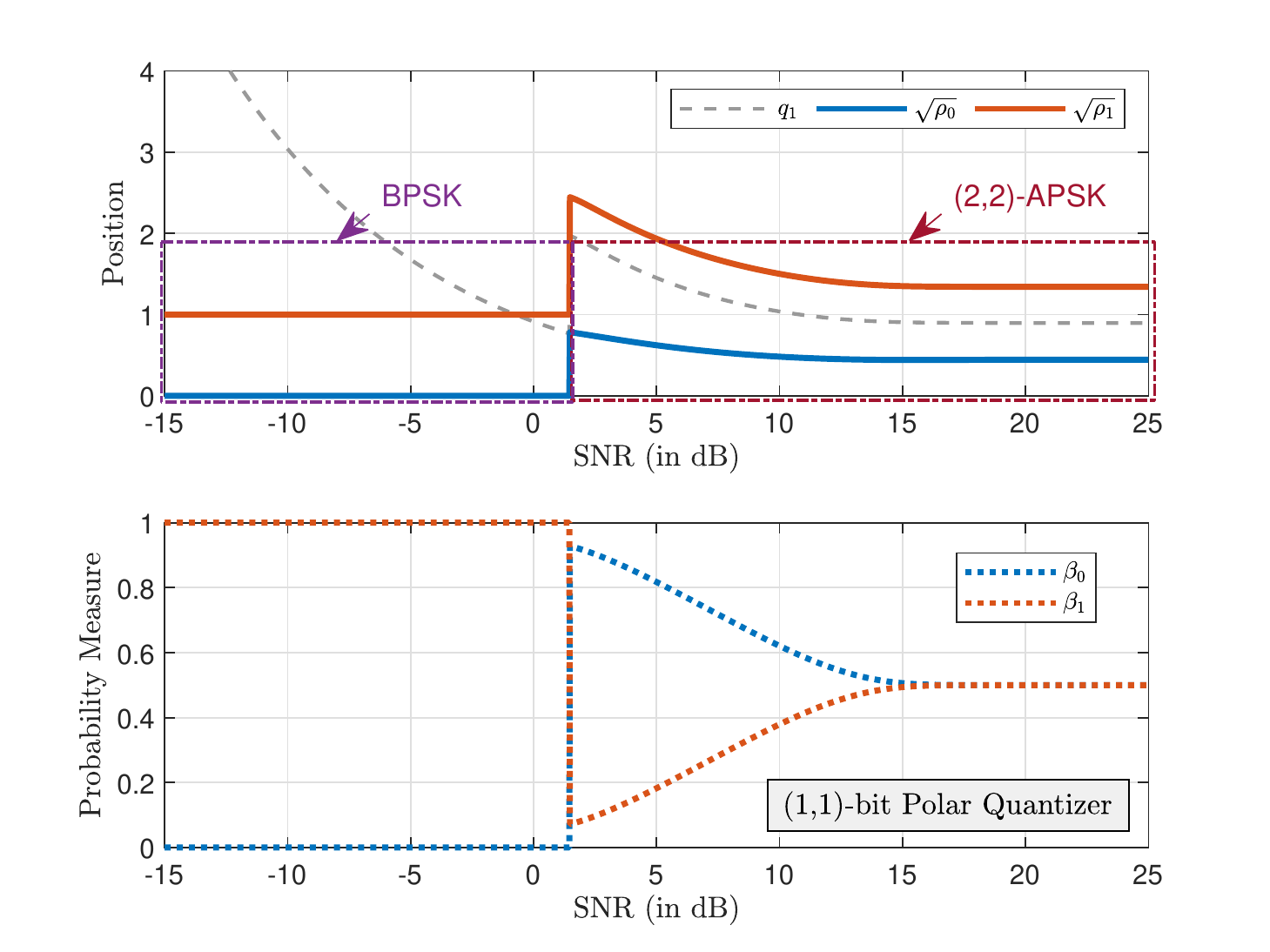}
    \label{fig:optimal_input_1_1}
} 
     \caption{Capacity-achieving input and optimal quantizer for AWGN channel with $(b_1,1)$-bit polar quantizer at the output. The top plot depicts the optimal position of the amplitude levels and $q_1$ as a function of SNR and the bottom plot shows the respective probabilities of these amplitude levels. The phase quantization bits are set as follows: (a) $b_1 = 4$, (b) $b_1 = 3$, (c) $b_1 = 2$, (d) $b_1 = 1$.}
     \label{fig:optimal_input_set}
\end{figure*}

Next, we investigate the parameters of the capacity-achieving input and the optimal position of the magnitude quantizer threshold $q_1$. The top plot of each subfigure of Figure \ref{fig:optimal_input_set} depicts the optimal locations of the amplitude levels, $\sqrt{\rho_0}$ and $\sqrt{\rho_1}$, and magnitude threshold, $q_1$, whereas the bottom plot of each subfigure of Figure \ref{fig:optimal_input_set} gives the corresponding probabilities of $\sqrt{\rho_0}$ and $\sqrt{\rho_1}$ (denote as $\beta_0$ and $\beta_1$, respectively). For a $(4,1)$-bit polar-quantized channel (see Figure \ref{fig:optimal_input_4_1}), the capacity-achieving input in the low SNR regime is 16-PSK. This is because the probability of $\sqrt{\rho_0}$ is zero so only one amplitude level is present in the optimal input distribution. At around 1.8 dB, an additional mass point starts to emerge at the origin (i.e. $\rho_0 = 0$ with $\beta_0 > 0$). At this point, the capacity-achieving input becomes an on-off 16-PSK modulation scheme with an off-state probability $\beta_0$. The off-state probability gradually increases as SNR is increased. However, at 5.25 dB, a threshold effect is noticed in which a sharp transition in the optimal parameter values occurs. More precisely, the SNR is high enough such that two non-zero amplitude levels can be reliably distinguished by the polar-quantized receiver. The capacity-achieving input shifts from an on-off 16-PSK modulation scheme to a $(16,2)$-APSK modulation scheme. As SNR is increased further, the capacity-achieving input eventually converges to an equiprobable $(16,2)$-APSK constellation, with the midpoint of the amplitude levels coinciding with $q_1$.


The same trend is observed for the $(3,1)$-bit polar-quantized AWGN channel (see Figure \ref{fig:optimal_input_3_1}) except that the optimal input distribution has 8 phase values instead of 16. In this case, an 8-PSK achieves capacity in the low SNR regime and then transitions to an on-off 8-PSK when a certain SNR level is attained. As SNR is increased further, the capacity-achieving input shifts to an $(8,2)$-APSK scheme. At this point, one might expect that the capacity-achieving input of a $(b_1,1)$-bit polar-quantized AWGN evolves in a similar manner for any $b_1$ as SNR is increased. In fact, some parallels can be drawn between our numerical results in Figures \ref{fig:optimal_input_4_1} and \ref{fig:optimal_input_3_1} and the numerical results of Singh et al. \cite{Singh:2009} when they investigated the capacity-achieving input of real AWGN channel with 2-bit output quantization. In their study, they noticed that BPSK is optimal in the low SNR regime but a mass point at the origin eventually appears when SNR is increased to a certain value. When the channel is good enough such that four input mass points can be disambiguated, the capacity-achieving input distribution becomes a 4-ary pulse amplitude modulation (4-PAM). Is there always a region in between the low SNR regime (for which $2^{b_1}$-PSK is optimal) and the high SNR regime (for which $(2^{b_1},2)$-APSK is optimal) such that on-off keying input is capacity-achieving? Moreover, is PSK always capacity-achieving in the low SNR regime?

\begin{figure*}[t]
    \centering
    \includegraphics[width =.95\textwidth]{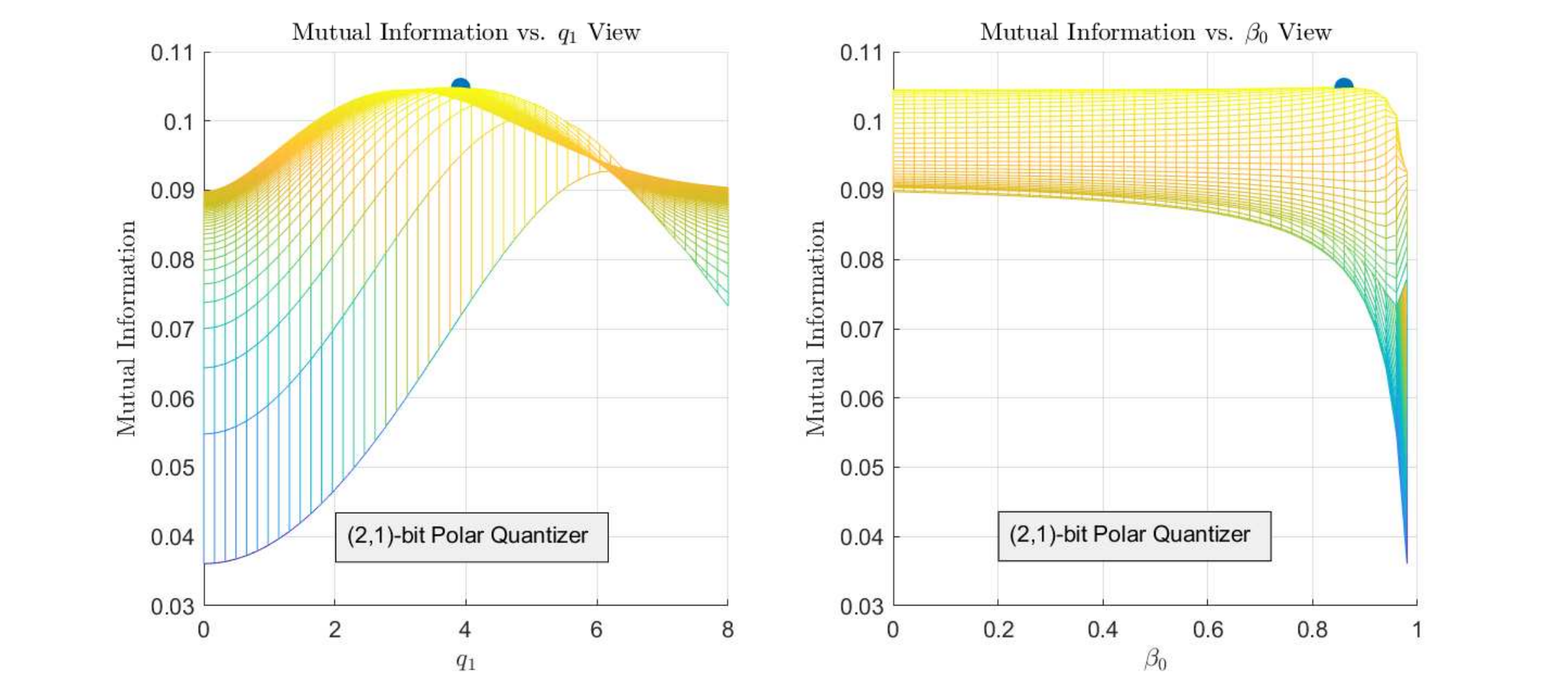}
    \caption{Mutual information (M.I.) surface vs $q_1$ and $\beta_0$ at $\mathrm{SNR} = -10\mathrm{ dB}$ for $b_1 = 2$. The lower amplitude is set to $\sqrt{\rho_0} = 0$ and the right and left plots depict the perspectives of M.I. vs $\beta_0$ and M.I. vs $q_1$, respectively. The blue circle marker indicates the maximum point of the surface plot.}
    \label{fig:capacity_surface_2_1}
\end{figure*}

\begin{figure*}[t]
  \centering
  \subfloat[]{
\includegraphics[width =.95\textwidth]{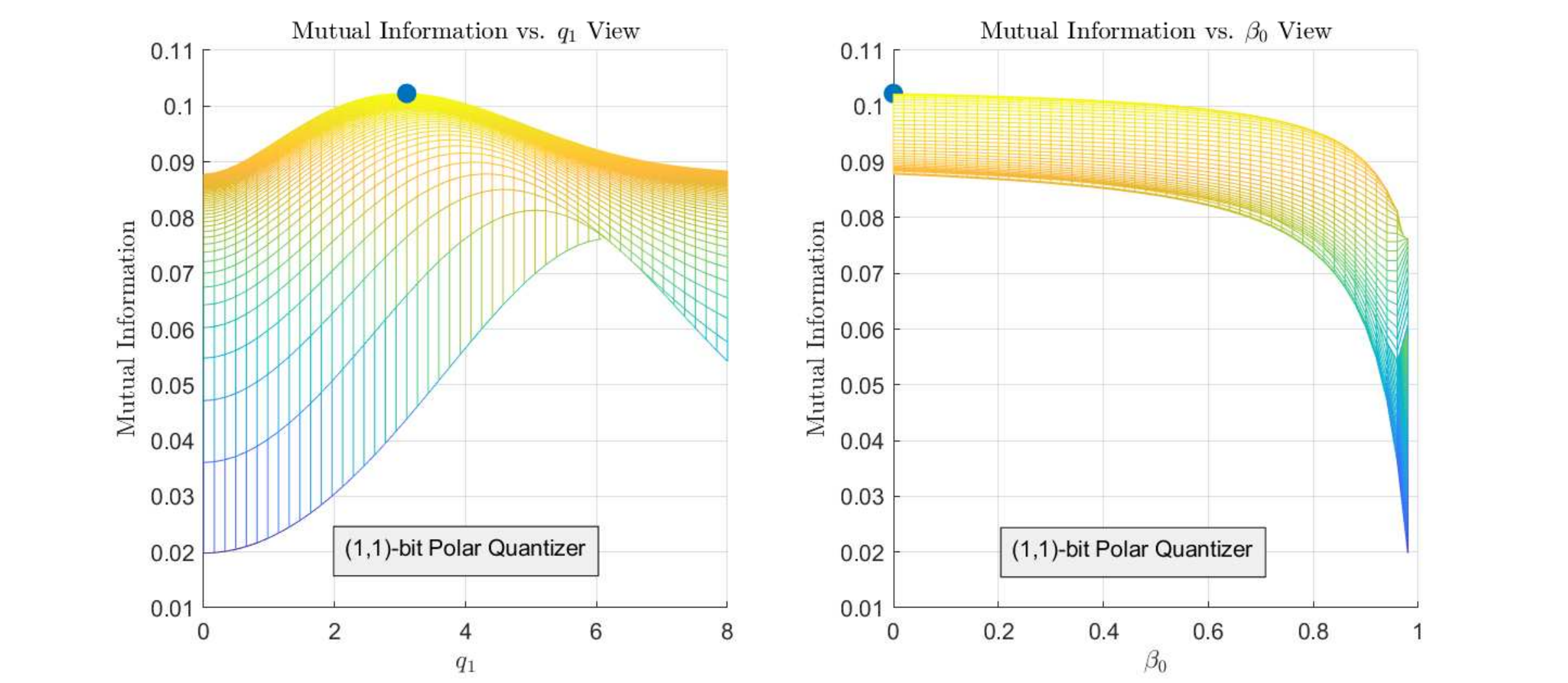}
\label{fig:capacity_surface_1_1}
}
\\
\vspace{-.25cm}
  \subfloat[]{
\includegraphics[width =.95\textwidth]{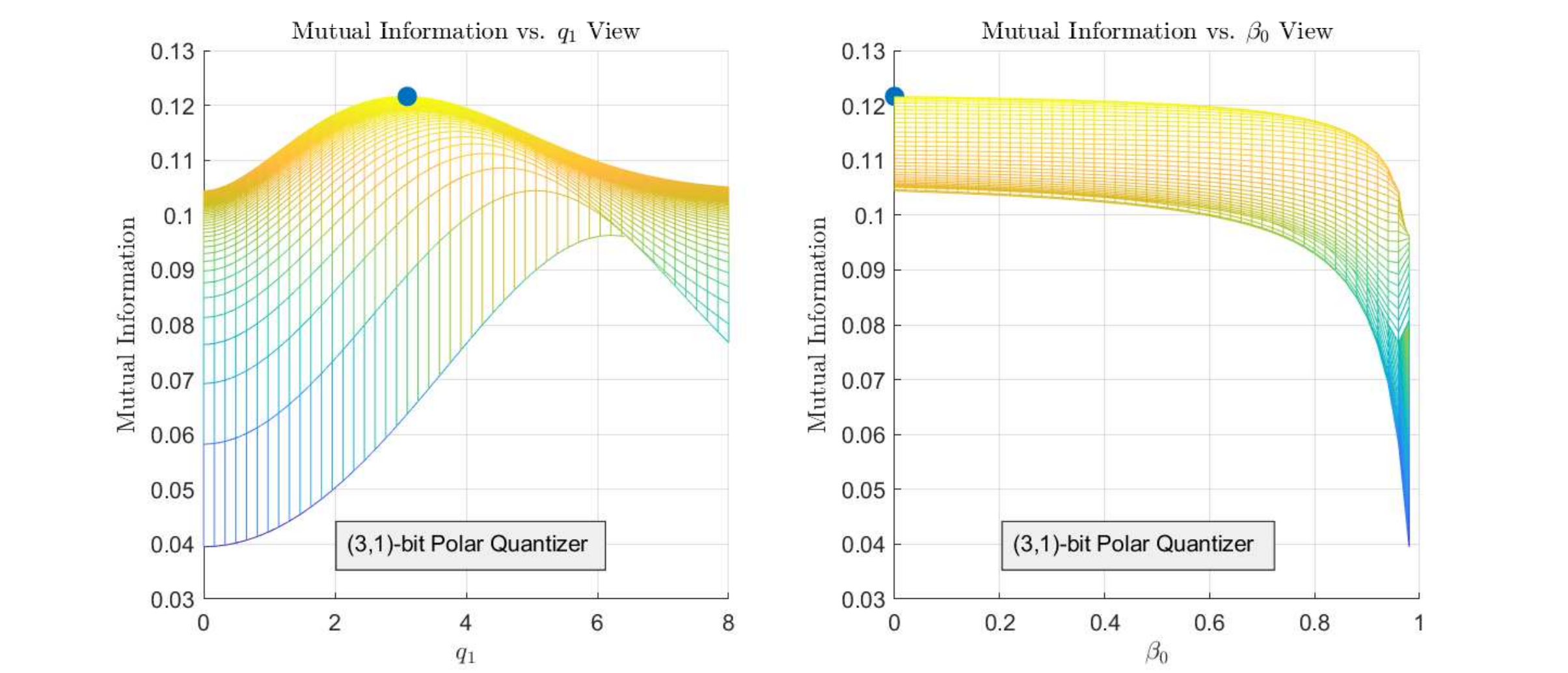}
\label{fig:capacity_surface_3_1}
}

     \caption{Mutual information (M.I.) surface plots against $q_1$ and $\beta_0$ at $\mathrm{SNR} = -10\mathrm{ dB}$. The lower amplitude is set to $\sqrt{\rho_0} = 0$ and the right and left plots depict the perspectives of M.I. vs $\beta_0$ and M.I. vs $q_1$, respectively. The blue circle marker indicates the maximum point of the surface plot. The phase quantization bits are set as follows: (a) $b_1 = 1$ and (b) $b_1 = 3$. }
     \label{fig:capacity_surface}
\end{figure*}

Our numerical results for the $(1,1)$-bit polar-quantized AWGN channel (see Figure \ref{fig:optimal_input_1_1}) suggest that a region where an on-off keying structure is optimal may not always exist in some configurations of the polar-quantized AWGN channel. Here, the BPSK, which is optimal in the low SNR regime, directly transitions to a $(2,2)$-APSK (or 4-PAM) when the SNR exceeds 1.45 dB. To address the second question, we turn our attention to the capacity-achieving input of $(2,1)$-bit polar-quantized channel (see Figure \ref{fig:optimal_input_2_1}). Here, we observed that the capacity-achieving input in the low SNR regime is an on-off QPSK scheme rather than QPSK. In addition, its off-state probability approaches unity as SNR is made arbitrarily small. To validate this peculiar observation in the low SNR regime of the $(2,1)$-bit polar-quantized AWGN channel, we plot the objective function in (\ref{eq:cap_numeric}) against $q_1$ and $\beta_0$ in Figure \ref{fig:capacity_surface_2_1}. The SNR is set to -10 dB and the lower amplitude level $\sqrt{\rho_0}$ is placed at the origin. The blue circle pinpoints the maximum value of the plot; thus showing that the capacity is achieved by an on-off keying QPSK with off-state probability $\beta_0 = 0.86$. Note, however, that the optimality of the on-off keying structure also relies on a specific choice of $q_1$ and that this choice for $q_1$ should grow unbounded for vanishing SNR. In case an upper bound on $q_1$ is imposed, the capacity-achieving input for this channel at vanishing SNR becomes QPSK. This observation has some resemblance to the results established by Koch et al. in \cite{Koch:2013} for 1-bit quantization in the low SNR regime. In their work, they showed that the low SNR capacity is achieved when an asymmetric 1-bit quantizer and an on-off keying input are used under an average power constraint, provided that the threshold of the quantizer is allowed to grow unbounded at vanishing SNR. Otherwise, BPSK input with symmetric output quantization is the optimal communication strategy.

The surface plots of the mutual information of $(1,1)$-bit and $(3,1)$-bit polar-quantized AWGN channels are given in Figures \ref{fig:capacity_surface_1_1} and \ref{fig:capacity_surface_3_1}, respectively. Since the blue circle is located at $\beta_0 = 0$, we verify that the capacity-achieving input for these channels at -10 dB are indeed BPSK and 8-PSK, respectively. An intriguing observation from these surface plots is that the capacity is achieved by a specific value of $q_1$ despite having no information encoded in the amplitude of the capacity-achieving input. One possible explanation for this is that the received samples falling above $q_1$ (i.e. $y_2 = 1$) can be tagged as ``unreliable" since they should have been corrupted by a large instantaneous additive noise in order to fall at this magnitude quantization region. This additional information can be exploited by the decoder to increase communication robustness against noise. Finally, we note that the capacity curves produced in Figure \ref{fig:cap_vs_snr} remain continuous despite the sharp transitions observed in the optimal values of $\rho_0$, $\beta_0$, and $q_1$ as SNR is varied.

\subsection{Practical Implications}

One practical advantage of channels with polar quantization over channels with I/Q quantization is having a more detailed description of the capacity-achieving input structure. Because of the established results on the phase components of the optimal input, the complexity of the optimization problem would only scale with the number of magnitude quantization bits. For instance, in a $(b_1,b_2)$-bit polar-quantized AWGN channel, there will be at most $2^{b_2}$ pairs of amplitude levels and probability values needed to be identified numerically. On the other hand, numerical approaches for AWGN channel with $b$-bit I/Q quantization would need to find at most $2^{2b}$ complex-valued mass points and their $2^{2b}-1$ respective probability masses, as illustrated in \cite{Vu:2019}.

Aside from reducing the complexity of the optimization problem, the structure of the capacity-achieving input has an added benefit of being more robust against nonlinear amplifier distortion compared to conventional QAM schemes \cite{Thomas:1974}. This is due to the ``concentric rings" structure of APSK which minimizes the amplitude variations of the transmitted signal. Consequently, this results in a lower peak-to-average power ratio (PAPR) as compared to that of QAM schemes\footnote{We exclude the on-off QPSK scheme with $\beta_0$ approaching unity in the low SNR regime of $(2,1)$-bit polar-quantized channel since its PAPR$\rightarrow\infty$.} \cite{Baldi:2012}.

\section{Conclusion}\label{section-conclusion}

In this work, we extend the capacity results of our previous works \cite{bernardo2021phase,bernardo2021TIT} to AWGN channel with polar quantization at the output. Our first contribution is a rigorous proof showing that either a $(2^{b_1},L)$-APSK scheme with $L \leq 2^{b_2}$ or an on-off $(2^{b_1},L')$-APSK scheme with $L' \leq 2^{b_2}-1$ is the capacity-achieving input distribution for a $(b_1,b_2)$-bit polar-quantized AWGN channel. We also show that the angles of the optimal input mass points can be derived analytically. Thus, the dimension of the optimization problem does not scale with $b_1$. We also note that Theorem \ref{theorem:AWGN_case} simplifies to \cite[Theorem 1]{bernardo2021TIT} when $b_2 = 0$ (i.e. no magnitude quantizer branch at the receiver). The derived capacity results also extend to Gaussian MISO channel with polar quantization at the output.

By leveraging on this analytical result, we evaluate the capacity of $(b_1,1)$-bit polar-quantized AWGN channels with numerically-optimized magnitude quantizer as well as the input distribution that achieves the capacity. We show that a suboptimal choice of $q_1$ can still achieve near-optimal performance in the low SNR regime but the choice of $q_1$ becomes more crucial in the high SNR regime. A threshold effect is observed at different SNR values at which sharp changes in the optimal parameters of the capacity-achieving input occur. More precisely, a sufficiently small increment at these SNR points increases the number of amplitude levels of the optimal input. The number of phase quantization bits also affects how the structure of the capacity-achieving input in some SNR regimes. For instance, the capacity-achieving inputs for the polar quantizers considered in Section \ref{section-numerical_analysis} have a PSK structure in the low SNR regime except for the $(2,1)$-bit polar quantizer. An analytical explanation for this odd observation is left as an open problem. 

An important direction for future research is to design computationally-efficient optimization methods for larger $b_2$ that exploit the properties of the capacity-achieving input. Such optimization method would enable an accurate capacity evaluation for $(b_1,b_2)$-bit polar-quantized channels with arbitrary $b_1$ and $b_2$; thus giving a wider perspective on how a specific polar quantization configuration impacts the optimal APSK structure. It is also of interest to generalize this result to other types of channels with polar quantization at the output. Can we prove that APSK achieves ergodic capacity in the presence of fading? How does the knowledge of fading state impact the structure of the optimal input? Lastly, while characterization of the capacity limits of polar-quantized channels is a fundamental step towards advancing communication systems with low-precision polar quantization, the design of other receiver functionalities (e.g. timing recovery, gain
control, channel estimation) for ADC-constrained polar receivers is also essential and worth exploring.

\begin{appendices}
\section{Proof of Proposition \ref{proposition:polar_symmetry_distribution}}\label{proof_symmetry}

We first define the notations
\begingroup
\allowdisplaybreaks
\begin{align*}
         &H_{F_U}(Y_1,Y_2)\\
         &\quad= -\int_{\mathbb{C}}\sum_{y_1=0}^{2^{b_1}-1}\sum_{y_2=0}^{2^{b_2}-1}W_{y_1,y_2}^{(b_1)}(u)\log p(y_1,y_2;F_U)\;dF_U\\
         &H_{F_U}(Y_1,Y_2|U)\\
         &\quad= -\int_{\mathbb{C}}\sum_{y_1=0}^{2^{b_1}-1}\sum_{y_2=0}^{2^{b_2}-1}W_{y_1,y_2}^{(b_1)}(u)\log W_{y_1,y_2}^{(b_1)}(u)\;dF_U,
\end{align*}%
\endgroup
where we used the subscript $F_U$ to note that the entropy and conditional entropy are induced by the input distribution in the subscript. We want to show that
\begin{align*}
    &H_{F_U^{s}}(Y_1,Y_2) - H_{F_U^{s}}(Y_1,Y_2|U) \\
    &\qquad\qquad\qquad\qquad\geq H_{F_U}(Y_1,Y_2) - H_{F_U}(Y_1,Y_2|U)
\end{align*}
holds for any distribution $F_U$. The conditional output entropy $H(Y_1,Y_2|U)$ using $F_U^s$ can be expressed as
\begingroup
\allowdisplaybreaks
\begin{align*}
    &H_{F_U^s}(Y_1,Y_2|U)\\
    &\quad= -\int_{\mathbb{C}}\sum_{y_1=0}^{2^{b_1}-1}\sum_{y_2=0}^{2^{b_2}-1}W_{y_1,y_2}^{(b_1)}(u)\log W_{y_1,y_2}^{(b_1)}(u)\\
    &\quad\quad\quad\cdot d\left[\frac{1}{2^{b_1}}\sum_{i = 0}^{2^{b_1}-1}F_U(ue^{j\frac{2\pi i}{2^{b_1}}})\right]\\
    &\quad=-\int_{\mathbb{C}}\sum_{y_1=0}^{2^{b_1}-1}\sum_{y_2=0}^{2^{b_2}-1}\Bigg\{\frac{1}{2^{b_1}}\sum_{i = 0}^{2^{b_1}-1}W_{y_1,y_2}^{(b_1)}(ue^{-j\frac{2\pi i}{2^{b_1}}})\\
    &\quad\qquad\qquad\cdot\log W_{y_1,y_2}^{(b_1)}(ue^{-j\frac{2\pi i}{2^{b_1}}})\Bigg\} \;dF_U.
\end{align*}
\endgroup
Due to the circular structure of the phase quantizer output $Y_1$ and Lemma \ref{lemma:symmetry_Wy}, we have
\begin{align*}
    &\frac{1}{2^{b_1}}\sum_{i = 0}^{2^{b_1}-1}W_{y_1,y_2}^{(b_1)}(ue^{-j\frac{2\pi i}{2^{b_1}}})\log W_{y_1,y_2}^{(b_1)}(ue^{-j\frac{2\pi i}{2^{b_1}}}) \\
    &\qquad\qquad= \frac{1}{2^{b_1}}\sum_{i = 0}^{2^{b_1}-1}W_{y_1,y_2}^{(b_1)}(u)\log W_{y_1,y_2}^{(b_1)}(u)\\
    &\qquad\qquad= W_{y_1,y_2}^{(b_1)}(u)\log W_{y_1,y_2}^{(b_1)}(u).
\end{align*}
The last line follows from the fact that the summation term in the previous line does not depend on $i$. Consequently, $H_{F_U^s}(Y|U) = H_{F_U}(Y|U)$. To prove the claim, we need to show that $H_{F_U^s}(Y_1,Y_2) \geq H_{F_U}(Y_1,Y_2)$. The output PMF $p(y_1,y_2;F_U^s)$ is
\begingroup\allowdisplaybreaks
\begin{align*}
      p(y_1,y_2;F_U^s) =&\int_{\mathbb{C}} W_{y_1,y_2}^{(b_1)}(u)\;d\left[\frac{1}{2^{b_1}}\sum_{i = 0}^{2^{b_1}-1}F_U(ue^{j\frac{2\pi i}{2^{b_1}}})\right]\\
      =&\int_{\mathbb{C}}\left[\frac{1}{2^{b_1}}\sum_{i = 0}^{2^{b_1}-1} W_{y_1,y_2}^{(b_1)}\left(ue^{-j\frac{2\pi i}{2^{b_1}}}\right)\right]\;dF_U\\
      =&\int_{\mathbb{C}}\left[\frac{1}{2^{b_1}}\sum_{i = 0}^{2^{b_1}-1} W_{y_1+i,y_2}^{(b_1)}\left(u\right)\right]\;dF_U\\
      =&\frac{1}{2^{b_1}}\int_{\mathbb{C}}W_{y_2}\left(u\right)\;dF_U.
\end{align*}
\endgroup
The second equality follows from rotating $U$. The third equality follows from Lemma \ref{lemma:symmetry_Wy}. Finally, we introduce the function
\begin{align}
    W_{y_2}(u) = \sum_{y_1 = 0}^{2^{b_1}-1}p(y_1,y_2|u) = p(y_2|u)
\end{align}
in the last line. This $p(y_2|u)$ is equal to $V_{y_2}(t,\sigma/\sqrt{2})$ defined in (\ref{eq:V_y2}), which is invariant of $\theta$. Thus, we can simply use $p(y_2|\alpha)$. Without loss of generality, we write the output PMF as
\begin{align}\label{eq:p_y2_FA}
      p(y_1,y_2;F_U^s) =&\frac{1}{2^{b_1}}\int_{\mathbb{R}_+}p\left(y_2|\alpha\right)\;dF_{A} \nonumber\\
      =&\frac{1}{2^{b_1}}\cdot p(y_2;F_A),
\end{align}
where $p(y_2;F_A)$ is the marginal PMF of $Y_2$ induced by the choice of amplitude distribution $F_A$. Consequently, the output entropy becomes
\begin{align}
    H_{F_U^s}(Y_1,Y_2) =& \log 2^{b_1} - \sum_{y_2 = 0}^{2^{b_2}-1}p(y_2;F_A)\log p(y_2;F_A)\nonumber\\
    =& b_1 + H_{F_A}(Y_2),
\end{align}
which is maximized for some $F_A$ since $Y_1$ is uniformly distributed and is independent of $Y_2$.

\section{Proof of Lemma \ref{lemma:boundedness_input}}\label{proof_boundedness}

To prove boundedness of the support, we consider two cases of the KTC coefficient $\mu$.\\
\noindent \textbf{Case A ($\mu > 0$)}: \\
As $\alpha \rightarrow \infty$ for any $\theta_0 \in \mathcal{R}_{y'}^{\text{PH}}$, the conditional PMF $W_{y_1,y_2}^{(b_1)}\left(\frac{\alpha}{\sigma^2},\theta_0\right)$ converges to
\begin{align}\label{eq:limit_Wy_part1}
    \lim_{\alpha\rightarrow\infty}W_{y_1,y_2}^{(b_1)}\left(\frac{\alpha}{\sigma^2},\theta_0\right) = \mathbbm{1}_{\{(y',2^{b_2}-1)\}}\left(y_1,y_2\right)
\end{align}
when $\theta_0 \ne \frac{2\pi y'}{2^{b_1}}$ (i.e. when $\theta_0$ does not fall exactly at the boundary of $\mathcal{R}_{y'}^{\text{PH}}$), and 
\begin{align}\label{eq:limit_Wy_part2}
    &\lim_{\alpha\rightarrow\infty}W_{y_1,y_2}^{(b_1)}\left(\frac{\alpha}{\sigma^2},\theta_0\right) \nonumber\\
    &\qquad= \frac{1}{2}\mathbbm{1}_{\{(y',2^{b_2}-1),(y'-1,2^{b_2}-1)\}}\left(y_1,y_2\right)
\end{align}
when $\theta_0 = \frac{2\pi y'}{2^{b_1}}$ (i.e. when $\theta_0$ falls exactly at the boundary of $\mathcal{R}_{y'}^{\text{PH}}$). The notation $\mathbbm{1}_{A}(\cdot,*)$ refers to the indicator function; which is 1 if $(\cdot,*)\in A$ and 0 otherwise. This property, combined with the continuity of a discrete entropy function on its probability law, gives
\begin{align}
    &\lim_{\alpha\rightarrow\infty} d\left(\frac{\alpha}{\sigma^2},\theta_0;F_U\right) \nonumber \\
    &\qquad= \begin{cases}
        -\log p(y_2 = 2^{b_2}-1;F_U),& \text{if $\theta_0 \ne \frac{2\pi y'}{2^{b_1}}$}\\
        1-\log p(y_2 = 2^{b_2}-1;F_U),& \text{if $\theta_0 = \frac{2\pi y'}{2^{b_1}}$}.
    \end{cases}
\end{align}
Here, we used the alternative expression for $d(u;F_U)$ described in Footnote \ref{footnote:divergence}. Since $C$, $b_1$, and $\mu$ are non-negative numbers and $\underset{|u|^2\rightarrow\infty}{\lim} d(u;F_U)$ is finite, the LHS of (\ref{eq:KTC}) grows unbounded as $\alpha\rightarrow\infty$. Equivalently, equality in (\ref{eq:KTC}) is not achieved so $u\in F_U^*$ cannot have an unbounded magnitude.\\
\noindent \textbf{Case B ($\mu = 0$)}: \\
In this case, the KTC becomes $C  \geq b_1 + d(u;F_U^*)$. Similar to the approach in \cite{Rahman:2020}, we want to show that there exists a finite constant $\alpha_0$ such that for $\alpha > \alpha_0$, equality in (\ref{eq:KTC}) cannot be achieved with $\mu = 0$ and any $\theta_0 \in \mathcal{R}_{y'}^{\text{PH}}$. Mathematically,
\begin{align*}
&\exists \alpha_0\in\mathbb{R^+}\;|\;\forall \alpha > \alpha_0:\\
&\qquad\qquad d\left(\frac{\alpha}{\sigma^2},\theta_0;F_U^*\right) < \underset{\alpha'\rightarrow\infty}{\lim}\;  d\left(\frac{\alpha'}{\sigma^2},\theta_0;F_U^*\right).    
\end{align*}
Consider first $\theta_0 \ne \frac{2\pi y'}{2^{b_1}}$. Due to (\ref{eq:limit_Wy_part1}), it follows that there exists a constant $\alpha_1 \in \mathbb{R}_+$ such that
\begin{align*}
    W_{k,l}^{(b_1)}\left(\frac{\alpha_1}{\sigma^2},\theta_0\right) <&\; p(y_2 = l;F_U)
\end{align*}
for $(k,l) \ne (y',2^{b_2}-1)$, and
\begin{align*}
    W_{y',2^{b_2}-1}^{(b_1)}\left(\frac{\alpha_1}{\sigma^2},\theta_0\right) >&\; p(y_2 = 2^{b_2}-1;F_U)
\end{align*}
otherwise. Therefore, it also follows that
\begingroup
\allowdisplaybreaks
 \begin{align*}
    &d\left(\frac{\alpha}{\sigma^2},\theta_0;F_U^*\right)\\
    &\qquad=\; \sum_{y_1=0}^{2^{b_1}-1}\sum_{y_2=0}^{2^{b_2}-1}W_{y_1,y_2}^{(b_1)}\left(\frac{\alpha}{\sigma^2},\theta_0\right)\log \frac{W_{y_1,y_2}^{(b_1)}\left(\frac{\alpha}{\sigma^2},\theta_0\right)}{p(y_2;F_U)}\\
    &\qquad<\; W_{y',2^{b_2}-1}^{(b_1)}\left(\frac{\alpha}{\sigma^2},\theta_0\right)\log \frac{W_{y',2^{b_2}-1}^{(b_1)}\left(\frac{\alpha}{\sigma^2},\theta_0\right)}{p(y_2 = 2^{b_2}-1;F_U)}\\
    &\qquad<\; -\log p(y_2 = 2^{b_2}-1;F_U) =\underset{\alpha\rightarrow\infty}{\lim}\;  d\left(\frac{\alpha}{\sigma^2},\theta_0;F_U^*\right).
\end{align*}
\endgroup
We do the same for $\theta_0 = \frac{2\pi y'}{2^{b_1}}$. Due to (\ref{eq:limit_Wy_part2}), it follows that there exists a constant $\alpha_2 \in \mathbb{R}_+$ such that
\begin{align*}
    W_{k,l}^{(b_1)}\left(\frac{\alpha_2}{\sigma^2},\theta_0\right) <&\; p(y_2 = l;F_U)
\end{align*}
for $(k,l) \notin \left\{(y',2^{b_2}-1),(y'-1,2^{b_2}-1)\right\}$. It then follows that
\begingroup
\allowdisplaybreaks
 \begin{align*}
     &d\left(\frac{\alpha}{\sigma^2},\theta_0;F_U^*\right)\\
     &\quad=\; \sum_{y_1=0}^{2^{b_1}-1}\sum_{y_2=0}^{2^{b_2}-1}W_{y_1,y_2}^{(b_1)}\left(\frac{\alpha}{\sigma^2},\theta_0\right)\log \frac{W_{y_1,y_2}^{(b_1)}\left(\frac{\alpha}{\sigma^2},\theta_0\right)}{p(y_2;F_U)}\\
    &\quad<\; W_{y',2^{b_2}-1}^{(b_1)}\left(\frac{\alpha}{\sigma^2},\theta_0\right)\log \frac{W_{y',2^{b_2}-1}^{(b_1)}\left(\frac{\alpha}{\sigma^2},\theta_0\right)}{p(y_2 = 2^{b_2}-1;F_U)} \\
    &\quad\quad\; + W_{y'-1,2^{b_2}-1}^{(b_1)}\left(\frac{\alpha}{\sigma^2},\theta_0\right)\log \frac{W_{y'-1,2^{b_2}-1}^{(b_1)}\left(\frac{\alpha}{\sigma^2},\theta_0\right)}{p(y_2 = 2^{b_2}-1;F_U)}\\
    &\quad<\; 1-\log p(y_2 = 2^{b_2}-1;F_U) =\; \underset{\alpha\rightarrow\infty}{\lim}\;  d\left(\frac{\alpha}{\sigma^2},\theta_0;F_U^*\right).
\end{align*}
\endgroup
Case B\ is established by setting $\alpha_0 = \max\{\alpha_1,\alpha_2\}$. Combining the results of both cases concludes the proof.

\section{Proof of Proposition \ref{proposition:discreteness_input}}\label{proof_discreteness}

First, let $P'_0 \leq P'$ and $R(y_1,y_2) = p(y_1,y_2;F_U^*)$ be the power and output distribution corresponding to the optimal input. Also, let $\mathcal{B}(l)$ be a Borel set of $x\in\mathbb{C}$ with $\alpha \leq l$. Due to Lemma \ref{lemma:boundedness_input}, there exists a finite $T$ such that $\text{supp}(F_U^*) \subset \mathcal{B}(T)$. Define a convex and compact set $\mathcal{S}$ to be
\begin{align*}
    \mathcal{S} = \{F_U| \text{supp}(F_U) \subset \mathcal{B}(T)\},
\end{align*}
 and the corresponding subset $\mathcal{M}$ of $\mathcal{S}$ as
\begin{align*}
    \mathcal{M} = \left\{F_U \in \mathcal{S}| p(y_1,y_2;F_U) = R(y_1,y_2)\right\}.
\end{align*}
It is clear that $F_U^* \in \mathcal{M}$ for some finite $T$ since the output PMF should be $p(y_1,y_2;F_U) = R(y_1,y_2)\;\forall y_1\in\{0,\cdots,2^{b_1}-1\},y_2\in\{0,\cdots,2^{b_2}-1\}$ and $F_U^*$ is bounded. Thus, we can rewrite the capacity formula as
\begin{align*}
        C  = & \underset{F_U\in\mathcal{M}}{\max}\;\left\{ I(F_U) - \mu\left(\int_{\mathbb{C}}|u|\;dF_U - P'\right)\right\}
\end{align*}
for some non-negative multiplier $\mu$. Note that $I(F_U) - \mu\left(\int_{\mathbb{C}}|u|\;dF_U - P'\right)$ is a linear functional of $F_U$. As such, it has a maximum at an extreme point in $\mathcal{M}$ and this extreme point is $F_U^*$. Moreover, we consider $\mathcal{M}$ as intersection of $\mathcal{S}$ and $2^{b_1+b_2}-1$ hyperplanes given by
\begin{equation*}
    \mathcal{H}_{y_1,y_2} : \int_{B(T)}W_{y_1,y_2}^{(b_1)}(u)\;dF_U = \frac{1}{2^b}
\end{equation*}
for all $(y_1,y_2) \ne (2^{b_1}-1,2^{b_2}-1)$ (defining $\mathcal{H}_{2^{b}-1,2^{b_2}-1}$ is redundant since the probability of all mass points should sum up to 1). Given this, we can apply Dubins' Theorem \cite{Dubins:1962} in the same way as how \cite[Section V-B]{Vu:2019} and \cite[Proposition 1]{Rahman2:2020} used it to prove the discreteness of the optimal distribution and set an upper bound on the number of mass points. The optimal distribution $F_U^*$ is a convex combination of at most $2^{b_1+b_2}$ extreme points of $\mathcal{M}$. These extreme points are the set of unit masses $\delta(u)$ for some $u\in \mathcal{B}(T)$. Thus, the capacity is achieved by a discrete input distribution with at most $2^{b_1+b_2}$ mass points.

\section{Proof of Proposition \ref{proposition:optimal_angles}}\label{proof_optimal_angles}

The expression for $\mathcal{L}\left(\alpha,\theta,\mu\right)$ can be explicitly written as
\begingroup
\allowdisplaybreaks
\begin{align}
     &\mathcal{L}\left(\alpha,\theta,\mu\right)\nonumber\\
     &\quad=C - b_1 + \mu(\alpha- P') \nonumber\\
     &\qquad- \sum_{y_1=0}^{2^{b_1}-1}\sum_{y_2=0}^{2^{b_2}-1}W_{y_1,y_2}^{(b_1)}\left(\frac{\alpha}{\sigma^2},\theta\right)\log \frac{W_{y_1,y_2}^{(b_1)}\left(\frac{\alpha}{\sigma^2},\theta\right)}{p(y_2;F_A^*)}\nonumber\\
     &\quad=C - b_1 + \mu(\alpha- P')\nonumber \\
     &\qquad+  \sum_{y_2=0}^{2^{b_2}-1}\underbrace{\sum_{y_1=0}^{2^{b_1}-1}W_{y_1,y_2}^{(b_1)}\left(\frac{\alpha}{\sigma^2},\theta\right)}_{V_{y_2}(\alpha)}\log p(y_2;F_A^*) \nonumber\\
     &\qquad- \sum_{y_1=0}^{2^{b_1}-1}\sum_{y_2=0}^{2^{b_2}-1}W_{y_1,y_2}^{(b_1)}\left(\frac{\alpha}{\sigma^2},\theta\right)\log W_{y_1,y_2}^{(b_1)}\left(\frac{\alpha}{\sigma^2},\theta\right),
\end{align}
\endgroup
where we note that $V_{y_2}(\alpha) = \sum_{y_1=0}^{2^{b_1}-1}W_{y_1,y_2}^{(b_1)}\left(\frac{\alpha}{\sigma^2},\theta\right)$ has been established in Appendix \ref{proof_symmetry}. As such, only the last summation term depends on $\theta$ so $\nabla_{\theta}\mathcal{L}\left(\alpha,\theta,\mu\right)$ becomes
\begingroup
\allowdisplaybreaks
\begin{align}\label{eq:diff_L_theta}
     &\nabla_{\theta}\mathcal{L}\left(\alpha,\theta,\mu\right) \nonumber\\
     &\quad= \frac{\partial}{\partial\theta}\left\{- \sum_{y_1=0}^{2^{b_1}-1}\sum_{y_2=0}^{2^{b_2}-1}W_{y_1,y_2}^{(b_1)}\left(\frac{\alpha}{\sigma^2},\theta\right)\log W_{y_1,y_2}^{(b_1)}\left(\frac{\alpha}{\sigma^2},\theta\right)\right\}\nonumber\\
     &\quad= - \sum_{y_1=0}^{2^{b_1}-1}\sum_{y_2=0}^{2^{b_2}-1}\nabla_{\theta}W_{y_1,y_2}^{(b_1)}\left(\frac{\alpha}{\sigma^2},\theta\right)\nonumber\\
     &\quad\qquad\qquad\cdot\left[1+\log W_{y_1,y_2}^{(b_1)}\left(\frac{\alpha}{\sigma^2},\theta\right)\right]\nonumber\\
     &\quad=\sum_{y_1=0}^{2^{b_1}-1}\sum_{y_2=0}^{2^{b_2}-1}\nabla_{\theta}W_{y_1,y_2}^{(b_1)}\left(\frac{\alpha}{\sigma^2},\theta\right)\log \frac{1}{W_{y_1,y_2}^{(b_1)}\left(\frac{\alpha}{\sigma^2},\theta\right)}\nonumber\\
     &\quad=\sum_{y_1=0,y_1\ne2^{b_1-1}}^{2^{b_1}-1}\sum_{y_2=0}^{2^{b_2}-1}\nabla_{\theta}W_{y_1,y_2}^{(b_1)}\left(\frac{\alpha}{\sigma^2},\theta\right)\nonumber\\
     &\quad\qquad\qquad\cdot\log \frac{W_{2^{b_1-1},y_2}^{(b_1)}\left(\frac{\alpha}{\sigma^2},\theta\right)}{W_{y_1,y_2}^{(b_1)}\left(\frac{\alpha}{\sigma^2},\theta\right)}.
\end{align}
\endgroup
The second line is obtained by applying chain rule of differentiation. Note that we have dropped the factor $\frac{1}{\ln 2}$ since it does not affect the sign of the differential. The third and last line follow from the fact that probabilities should sum up to 1. That is,
\begingroup
\allowdisplaybreaks
\begin{align*}
    \sum_{y_1=0}^{2^{b_1}-1}\sum_{y_2=0}^{2^{b_2}-1}W_{y_1,y_2}^{(b_1)}\left(\frac{\alpha}{\sigma^2},\theta\right) = 1. 
\end{align*}
\endgroup
Consequently, we get the following identities in terms of first order derivatives with respect to $\theta$: 
\begin{align*}
    \sum_{y_1=0}^{2^{b_1}-1}\sum_{y_2=0}^{2^{b_2}-1}\nabla_{\theta}W_{y_1,y_2}^{(b_1)}\left(\frac{\alpha}{\sigma^2},\theta\right) =& 0
\end{align*}
and
\begin{align*}
    &\sum_{y_2=0}^{2^{b_2}-1}\nabla_{\theta}W_{2^{b_1-1},y_2}^{(b_1)}\left(\frac{\alpha}{\sigma^2},\theta\right)\nonumber\\
    &\quad\qquad= - \sum_{y_1=0,y_1\ne2^{b_1-1}}^{2^{b_1}-1}\sum_{y_2=0}^{2^{b_2}-1}\nabla_{\theta}W_{y_1,y_2}^{(b_1)}\left(\frac{\alpha}{\sigma^2},\theta\right)
\end{align*}

An optimal angle should satisfy $\nabla_{\theta}\mathcal{L}\left(\alpha,\theta,\mu\right) = 0$. By some algebraic manipulation, we can rewrite (\ref{eq:diff_L_theta}) as \eqref{eq:diff_L_theta_2}. Now suppose we set $\theta = \frac{\pi}{2^{b_1}}$. Leibniz integral rule can be applied to get equations \eqref{subeq:diff_L_sym_pi_2b_a} - \eqref{subeq:diff_L_sym_pi_2b_c}. Note that $\tau\left(r,\phi,\nu\right)$ is even symmetric about $\phi = 0$. As such, we have
\begingroup
\allowdisplaybreaks
\begin{align*}
      \nabla_{\theta}W_{2^{b_1-1} - y_1,y_2}^{(b_1)}\left(\frac{\alpha}{\sigma^2},\frac{\pi}{2^{b_1}}\right) =& - \nabla_{\theta}W_{2^{b_1-1} + y_1,y_2}^{(b_1)}\left(\frac{\alpha}{\sigma^2},\frac{\pi}{2^{b_1}}\right)
\end{align*}
and
\begin{align*}
      \nabla_{\theta}W_{0,y_2}^{(b_1)}\left(\frac{\alpha}{\sigma^2},\frac{\pi}{2^{b_1}}\right) = 0.
\end{align*}
\endgroup
By combining this with Lemma \ref{lemma:specialvalues_Wy}.i, we get $\nabla_{\theta}\mathcal{L}\left(\alpha,\frac{\pi}{2^{b_1}},\mu\right) = 0$. Alternatively, we can write (\ref{eq:diff_L_theta}) as \eqref{eq:diff_L_theta_3}. Suppose we set $\theta = 0$. Then, the last term becomes zero due to Lemma \ref{lemma:specialvalues_Wy}.ii (the argument of $\log(\cdot)$ becomes 1). We apply Leibniz integral rule again to get equations \eqref{subeq:diff_L_sym_0_a} and \eqref{subeq:diff_L_sym_0_b}. Due to the even symmetry of $\tau\left(r,\phi,\nu\right)$ about $\phi = 0$, we have
\begingroup
\allowdisplaybreaks
\begin{align*}
      \nabla_{\theta}W_{2^{b_1-1} - y_1,y_2}^{(b_1)}\left(\frac{\alpha}{\sigma^2},0\right) =& - \nabla_{\theta}W_{2^{b_1-1}-1 + y_1,y_2}^{(b_1)}\left(\frac{\alpha}{\sigma^2},0\right).
\end{align*}
\endgroup
Combining this with Lemma \ref{lemma:specialvalues_Wy}.ii gives us $\nabla_{\theta}\mathcal{L}\left(\alpha,0,\mu\right) = 0$. Thus, the two stationary points within $\mathcal{R}_{2^{b_1-1}}^{\mathrm{PH}}$ occur at $\theta = 0$ (exactly at the phase quantization boundary) and $\theta = \frac{\pi}{2^{b_1}}$ (exactly at the middle of the phase quantization region). To prove that $\theta = 0$ is not a minimizer of $\mathcal{L}\left(\alpha,\theta,\mu\right)$, it suffices to show that
\begin{align*}
    \mathcal{L}\left(\alpha,\frac{\pi}{2^{b_1}},\mu\right)< \mathcal{L}\left(\alpha,0,\mu\right) \qquad\forall \alpha > 0,
\end{align*}
which, after some algebraic manipulation and using Lemma \ref{lemma:specialvalues_Wy}.i and Lemma \ref{lemma:specialvalues_Wy}.ii , simplifies to
\begin{align*}
    &\sum_{y_1=0}^{2^{b_1-1}-1}\sum_{y_2=0}^{2^{b_2}-1}\Bigg\{W_{y_1+1,y_2}^{(b_1)}\left(\frac{\alpha}{\sigma^2},\frac{\pi}{2^{b_1}}\right)\log W_{y_1+1,y_2}^{(b_1)}\left(\frac{\alpha}{\sigma^2},\frac{\pi}{2^{b_1}}\right)\\
    &\quad+W_{y_1,y_2}^{(b_1)}\left(\frac{\alpha}{\sigma^2},\frac{\pi}{2^{b_1}}\right)\log W_{y_1,y_2}^{(b_1)}\left(\frac{\alpha}{\sigma^2},\frac{\pi}{2^{b_1}}\right)\Bigg\}\\
    &\qquad<-2\sum_{y_1=0}^{2^{b_1-1}-1}\sum_{y_2=0}^{2^{b_2}-1}W_{y_1,y_2}^{(b_1)}\left(\frac{\alpha}{\sigma^2},0\right)\log W_{y_1,y_2}^{(b_1)}\left(\frac{\alpha}{\sigma^2},0\right),
\end{align*}
where
\[W_{y_1,y_2}^{(b_1)}\left(\frac{\alpha}{\sigma^2},0\right) \leq W_{y_1+1,y_2}^{(b_1)}\left(\frac{\alpha}{\sigma^2},\frac{\pi}{2^{b_1}}\right) \]
and
\[W_{y_1,y_2}^{(b_1)}\left(\frac{\alpha}{\sigma^2},0\right) \geq W_{y_1,y_2}^{(b_1)}\left(\frac{\alpha}{\sigma^2},\frac{\pi}{2^{b_1}}\right),
\]
for all $y\in\{0,\cdots,2^{b_1-1}-1\}$. Applying \cite[Lemma 8]{bernardo2021TIT} verifies the claim that $\theta = 0$ is not a minimizer of $\mathcal{L}\left(\alpha,\theta,\mu\right)$. The proof is completed by noting that $F_U^*$ should be a $\frac{2\pi}{2^{b_1}}$-symmetric distribution.
\begingroup
\allowdisplaybreaks
\begin{figure*}[!htb]
\begin{align}\label{eq:diff_L_theta_2}
    \nabla_{\theta}\mathcal{L}\left(\alpha,\theta,\mu\right) = & \sum_{y_2=0}^{2^{b_2}-1}\Bigg\{\sum_{y_1=1}^{2^{b_1-1}-1}\nabla_{\theta}W_{2^{b_1-1}-y_1,y_2}^{(b_1)}\left(\frac{\alpha}{\sigma^2},\theta\right)\log \frac{W_{2^{b_1-1},y_2}^{(b_1)}\left(\frac{\alpha}{\sigma^2},\theta\right)}{W_{2^{b_1-1} - y_1,y_2}^{(b)}\left(\frac{\alpha}{\sigma^2},\theta\right)}\nonumber\\
    &+\sum_{y_1=1}^{2^{b_1-1}-1}\nabla_{\theta}W_{2^{b_1-1}+y_1,y_2}^{(b_1)}\left(\frac{\alpha}{\sigma^2},\theta\right)\log \frac{W_{2^{b_1-1},y_2}^{(b_1)}\left(\frac{\alpha}{\sigma^2},\theta\right)}{W_{2^{b_1-1} + y_1,y_2}^{(b)}\left(\frac{\alpha}{\sigma^2},\theta\right)}+\nabla_{\theta}W_{0,y_2}^{(b_1)}\left(\frac{\alpha}{\sigma^2},\theta\right)\log \frac{W_{2^{b_1-1},y_2}^{(b_1)}\left(\frac{\alpha}{\sigma^2},\theta\right)}{W_{0,y_2}^{(b)}\left(\frac{\alpha}{\sigma^2},\theta\right)}\Bigg\}.
\end{align}
\hrulefill
\end{figure*}
\begin{figure*}[!htb]
\begin{align}
    \nabla_\theta W_{2^{b_1-1} - y_1,y_2}^{(b_1)}\left(\frac{\alpha}{\sigma^2},\frac{\pi}{2^{b_1}}\right)=&\left[\tau\left(\frac{q_{y_2+1}^2}{\sigma^2},-\frac{2\pi(y_1+0.5)}{2^{b_1}},\frac{\alpha}{\sigma^2}\right) -  \tau\left(\frac{q_{y_2}^2}{\sigma^2},-\frac{2\pi(y_1+0.5)}{2^{b_1}},\frac{\alpha}{\sigma^2}\right)\right]\nonumber\\
    &-\left[\tau\left(\frac{q_{y_2+1}^2}{\sigma^2},-\frac{2\pi(y_1-0.5)}{2^{b_1}},\frac{\alpha}{\sigma^2}\right) -  \tau\left(\frac{q_{y_2}^2}{\sigma^2},-\frac{2\pi(y_1-0.5)}{2^{b_1}},\frac{\alpha}{\sigma^2}\right)\right]
\label{subeq:diff_L_sym_pi_2b_a}\\
    \nabla_\theta W_{2^{b_1-1} + y_1,y_2}^{(b_1)}\left(\frac{\alpha}{\sigma^2},\frac{\pi}{2^{b_1}}\right)=&\left[\tau\left(\frac{q_{y_2+1}^2}{\sigma^2},\frac{2\pi(y_1-0.5)}{2^{b_1}},\frac{\alpha}{\sigma^2}\right) -  \tau\left(\frac{q_{y_2}^2}{\sigma^2},\frac{2\pi(y_1-0.5)}{2^{b_1}},\frac{\alpha}{\sigma^2}\right)\right]\nonumber\\
    &-\left[\tau\left(\frac{q_{y_2+1}^2}{\sigma^2},\frac{2\pi(y_1+0.5)}{2^{b_1}},\frac{\alpha}{\sigma^2}\right) -  \tau\left(\frac{q_{y_2}^2}{\sigma^2},\frac{2\pi(y_1+0.5)}{2^{b_1}},\frac{\alpha}{\sigma^2}\right)\right]\label{subeq:diff_L_sym_pi_2b_b}\\
    \nabla_\theta W_{0,y_2}^{(b_1)}\left(\frac{\alpha}{\sigma^2},\frac{\pi}{2^{b_1}}\right)=&\left[\tau\left(\frac{q_{y_2+1}^2}{\sigma^2},-\frac{\pi}{2^{b_1}},\frac{\alpha}{\sigma^2}\right) -  \tau\left(\frac{q_{y_2}^2}{\sigma^2},-\frac{\pi}{2^{b_1}},\frac{\alpha}{\sigma^2}\right)\right]\nonumber\\
    &-\left[\tau\left(\frac{q_{y_2+1}^2}{\sigma^2},\frac{\pi}{2^{b_1}},\frac{\alpha}{\sigma^2}\right) -  \tau\left(\frac{q_{y_2}^2}{\sigma^2},\frac{\pi}{2^{b_1}},\frac{\alpha}{\sigma^2}\right)\right]\label{subeq:diff_L_sym_pi_2b_c}
\end{align}
\hrulefill
\end{figure*}
\begin{figure*}[!htb]
\begin{align}\label{eq:diff_L_theta_3}
    \nabla_{\theta}\mathcal{L}\left(\alpha,\theta,\mu\right) = & \sum_{y_2=0}^{2^{b_2}-1}\Bigg\{\sum_{y_1=1}^{2^{b_1-1}-1}\nabla_{\theta}W_{2^{b_1-1}-y_1,y_2}^{(b_1)}\left(\frac{\alpha}{\sigma^2},\theta\right)\log \frac{W_{2^{b_1-1},y_2}^{(b_1)}\left(\frac{\alpha}{\sigma^2},\theta\right)}{W_{2^{b_1-1} - y_1,y_2}^{(b)}\left(\frac{\alpha}{\sigma^2},\theta\right)}\nonumber\\
    &\qquad+\sum_{y_1=1}^{2^{b_1-1}-1}\nabla_{\theta}W_{2^{b_1-1}-1+y_1,y_2}^{(b_1)}\left(\frac{\alpha}{\sigma^2},\theta\right)\log \frac{W_{2^{b_1-1},y_2}^{(b_1)}\left(\frac{\alpha}{\sigma^2},\theta\right)}{W_{2^{b_1-1}-1 + y_1,y_2}^{(b)}\left(\frac{\alpha}{\sigma^2},\theta\right)}\nonumber\\
    &\qquad+\nabla_{\theta}W_{2^{b_1-1}-1,y_2}^{(b_1)}\left(\frac{\alpha}{\sigma^2},\theta\right)\log \frac{W_{2^{b_1-1},y_2}^{(b_1)}\left(\frac{\alpha}{\sigma^2},\theta\right)}{W_{2^{b_1-1}-1,y_2}^{(b)}\left(\frac{\alpha}{\sigma^2},\theta\right)}\Bigg\}.
\end{align}
\hrulefill
\end{figure*}
\begin{figure*}[!htb]
\begin{align}
    \nabla_\theta W_{2^{b_1-1} - y_1,y_2}^{(b_1)}\left(\frac{\alpha}{\sigma^2},0\right)=&\left[\tau\left(\frac{q_{y_2+1}^2}{\sigma^2},-\frac{2\pi y_1}{2^{b_1}},\frac{\alpha}{\sigma^2}\right) -  \tau\left(\frac{q_{y_2}^2}{\sigma^2},-\frac{2\pi y_1}{2^{b_1}},\frac{\alpha}{\sigma^2}\right)\right]\nonumber\\
    &-\left[\tau\left(\frac{q_{y_2+1}^2}{\sigma^2},-\frac{2\pi(y_1-1)}{2^{b_1}},\frac{\alpha}{\sigma^2}\right) -  \tau\left(\frac{q_{y_2}^2}{\sigma^2},-\frac{2\pi(y_1-1)}{2^{b_1}},\frac{\alpha}{\sigma^2}\right)\right]\label{subeq:diff_L_sym_0_a}\\
    \nabla_\theta W_{2^{b_1-1} + y_1,y_2}^{(b_1)}\left(\frac{\alpha}{\sigma^2},0\right)=&\left[\tau\left(\frac{q_{y_2+1}^2}{\sigma^2},\frac{2\pi(y_1-1)}{2^{b_1}},\frac{\alpha}{\sigma^2}\right) -  \tau\left(\frac{q_{y_2}^2}{\sigma^2},\frac{2\pi(y_1-1)}{2^{b_1}},\frac{\alpha}{\sigma^2}\right)\right]\nonumber\\
    &-\left[\tau\left(\frac{q_{y_2+1}^2}{\sigma^2},\frac{2\pi y_1}{2^{b_1}},\frac{\alpha}{\sigma^2}\right) -  \tau\left(\frac{q_{y_2}^2}{\sigma^2},\frac{2\pi y_1}{2^{b_1}},\frac{\alpha}{\sigma^2}\right)\right]\label{subeq:diff_L_sym_0_b}
\end{align}
\hrulefill
\end{figure*}
\endgroup

\end{appendices}

\ifCLASSOPTIONcaptionsoff
  \newpage
\fi

\bibliographystyle{ieeetr}
\bibliography{references}

\newpage
\begin{IEEEbiography}[{\includegraphics[width=1in,height=1.25in,clip,keepaspectratio]{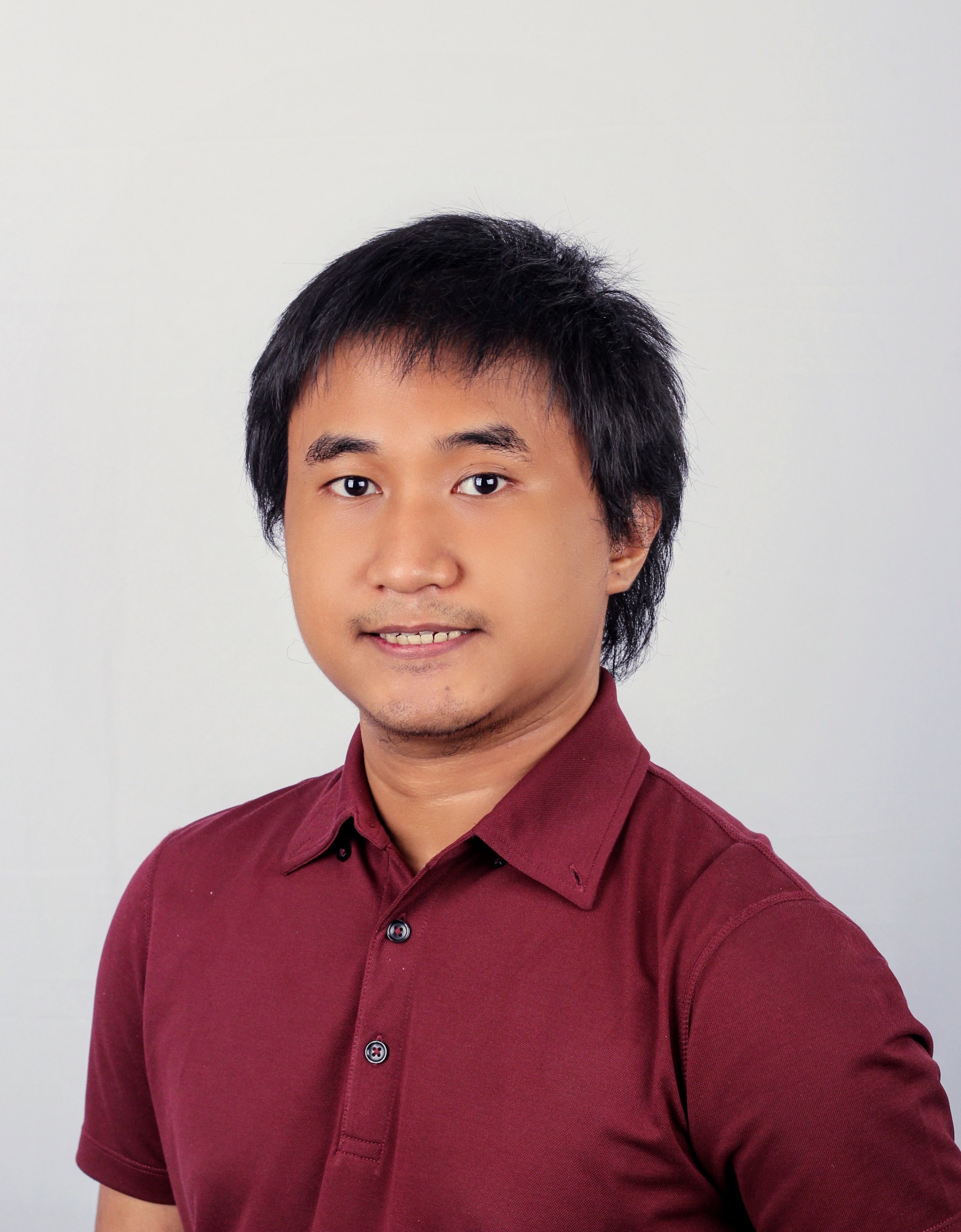}}]{Neil Irwin Bernardo} received his B.S. degree in Electronics and Communications Engineering from the University of the Philippines Diliman in 2014 and his M.S. degree in Electrical Engineering from the same university in 2016. He has been a faculty member of the University of the Philippines Diliman since 2014, and is currently on study leave to pursue a Ph.D. degree in Engineering at the University of Melbourne, Australia. His research interests include wireless communications, signal processing, and information theory.
\end{IEEEbiography}
\vskip 0pt plus -1fil
\begin{IEEEbiography}[{\includegraphics[width=1in,height=1.25in,clip,keepaspectratio]{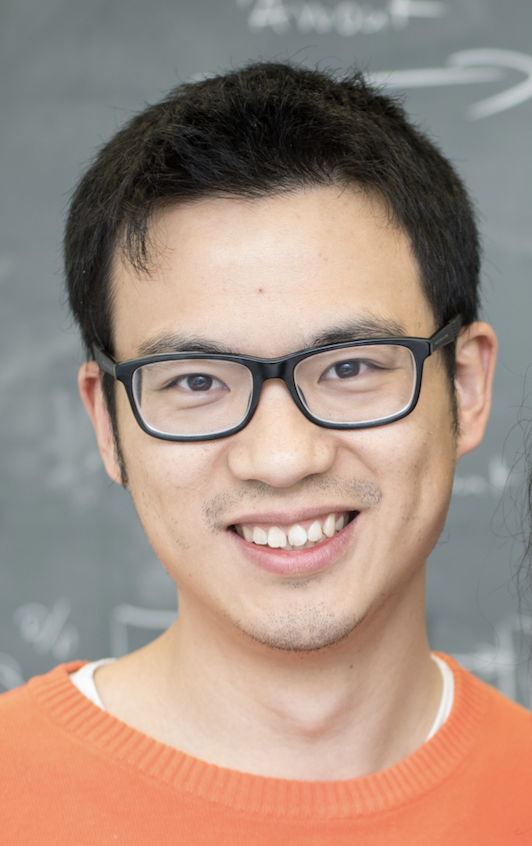}}]{Jingge Zhu} received the B.S. degree and M.S. degree in electrical engineering from Shanghai Jiao Tong University, Shanghai, China, in 2008 and 2011, respectively, the Dipl.-Ing. degree in technische Informatik from Technische Universit\"{a}t Berlin, Berlin, Germany in 2011 and the Doctorat \`{e}s Sciences degree from the Ecole Polytechnique F\'{e}d\'{e}rale (EPFL), Lausanne, Switzerland, in 2016. He was a post-doctoral researcher at the University of California, Berkeley from 2016 to 2018.  He is now a lecturer at the University of Melbourne, Australia. His research interests include information theory with applications in communication systems and machine learning. 

Dr. Zhu received the Discovery Early Career Research Award (DECRA) from the Australian Research Council in 2021, the IEEE Heinrich Hertz Award for Best Communications Letters in 2013, the Early Postdoc. Mobility Fellowship from Swiss National Science Foundation in 2015, and the Chinese Government Award for Outstanding Students Abroad in 2016.
\end{IEEEbiography}
\vskip 0pt plus -1fil
\begin{IEEEbiography}[{\includegraphics[width=1in,height=1.25in,clip,keepaspectratio]{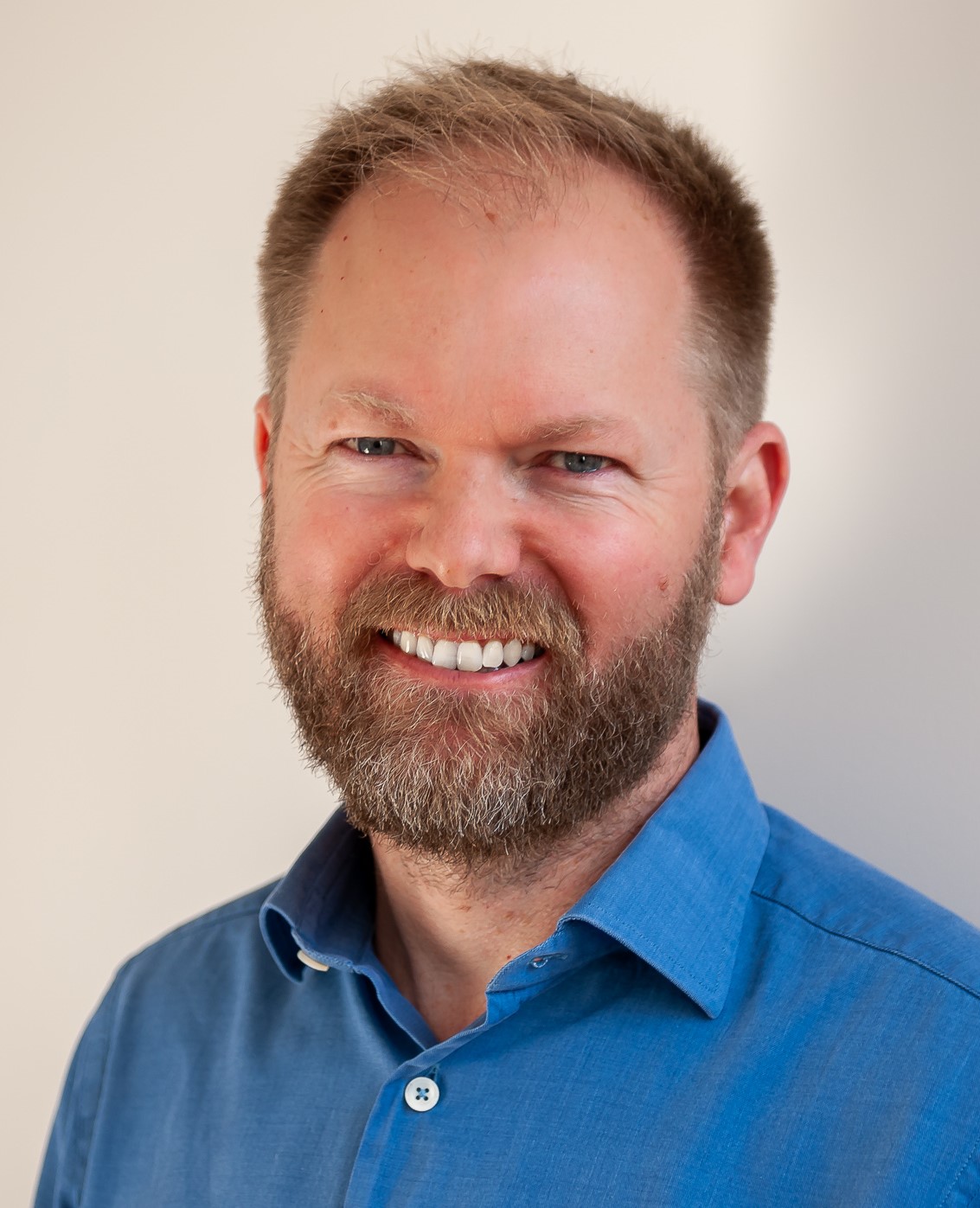}}]{Jamie Evans} was born in Newcastle, Australia, in 1970. He received the B.S. degree in physics and the B.E. degree in computer engineering from the University of Newcastle, in 1992 and 1993, respectively, where he received the University Medal upon graduation. He received the M.S. and the Ph.D. degrees from the University of Melbourne, Australia, in 1996 and 1998, respectively, both in electrical engineering, and was awarded the Chancellor's Prize for excellence for his Ph.D. thesis. From March 1998 to June 1999, he was a Visiting Researcher in the Department of Electrical Engineering and Computer Science, University of California, Berkeley. Since returning to Australia in July 1999 he has held academic positions at the University of Sydney, the University of Melbourne and Monash University. He is currently a Professor of Electrical and Electronic Engineering and Pro Vice-Chancellor (Education) at the University of Melbourne. His research interests are in communications theory, information theory, and statistical signal processing with a focus on wireless communications networks.
\end{IEEEbiography}

\end{document}